\documentclass[12pt,titlepage]{article}
\usepackage{axodraw,epsfig}

\def\_{\rule{.3em}{.15ex}}

\newcommand{\be}{\begin{equation}}
\newcommand{\ee}{\end{equation}}
\newcommand{\bea}{\begin{eqnarray}}
\newcommand{\eea}{\end{eqnarray}}
\newcommand{\f}{\frac}
\def\slash#1{\setbox0=\hbox{$#1$}#1\hskip-\wd0\dimen0=5pt\advance
       \dimen0 by-\ht0\advance\dimen0 by\dp0\lower0.5\dimen0\hbox
         to\wd0{\hss\sl/\/\hss}}

\def\ra{\rightarrow}

\def\TeV{\mathrm{TeV}}
\def\GeV{\mathrm{GeV}}

\begin{document}

\begin{titlepage}

\begin{flushright}
  {\bf hep-ph/0012031}\\
  {\bf TUM-HEP-395/00}\\[3mm]
  {\bf December 2000}
\end{flushright}

\vspace{2cm}

\begin{center}
\setlength {\baselineskip}{0.3in}

{\large \bf Chargino searches at LEP for complex MSSM parameters%
\footnote{This work was supported in part by the Polish
Committee for Scientific Research under the grant number 2~P03B~052~16
and by the French-Polish exchange program POLONIUM project 01391TB.}
}\\[10mm]

\setlength{\baselineskip}{0.2in}

{Nabil Ghodbane$^a$, Stavros Katsanevas$^a$, Imad Laktineh$^a$ and
Janusz Rosiek$^{b,}$\footnote{On leave of absence from Institute of
Theoretical Physics, Warsaw University, Ho{\.z}a 69, 00-681 Warsaw,
Poland}}

\vspace{10mm}

{\it $^a$Institut de Physique Nucl\'eaire de Lyon \\ 43 Bd du 11
novembre 1918, 69622 Villeurbanne cedex, France}\\[2mm]
{\it $^b$Physik Department, Technische Universit\"at M\"unchen\\
D-85748 Garching, Germany}

\vspace{2cm} 

{\bf Abstract}\\[5mm] 

\begin{minipage}{0.8\linewidth}
  {\small We reanalyze the results of the chargino searches at LEP (on
  the basis of the DELPHI 189 GeV data sample), including the
  possibility of complex MSSM parameters. We point out the possible
  differences between the complex- and real-parameter analysis. We
  check that the regions excluded by ``standard'' analyses remain
  generally robust against the introduction of complex parameters,
  with the exception of light sneutrino and low $\tan\beta$ scenario,
  where additional constraints like those given by the $Z^0$ width or
  electric dipole moment measurements are necessary.}
\end{minipage}

\end{center} 

\end{titlepage} 

\setlength{\baselineskip}{18pt}

\section{Introduction}

In the Minimal Supersymmetric Standard Model (MSSM) there are new
potential sources of CP non-conservation effects.  One can distinguish
two categories of such sources.  One is independent of the physics of
flavor non-conservation in the neutral current sector and the other is
closely related to it.  The first category is particularly interesting
from the point of view of collider experiments as it may affect the
direct searches for the supersymmetric Higgs bosons~\cite{WAGNER} and
other particles.  Complex phases may be present several
flavor-conserving parameters of the MSSM Lagrangian: $\mu$ parameter,
gaugino masses $M_i$, trilinear scalar couplings $A_i$ and soft Higgs
mixing term $m_{12}^2$. In principle they can be arbitrary (although
not all of them are physically independent).

Experimental constraints on the ``flavor-conserving'' phases come
mainly from the electric dipole moments of electron~\cite{EDM_E_EXP}
and neutron~\cite{EDM_N_EXP}:
\bea
E_e^{exp}<4.3\cdot 10^{-27} e\cdot cm\nonumber\\
E_n^{exp}<6.3\cdot 10^{-26} e\cdot cm\nonumber
\label{eq:edm_en_exp}
\eea

Until recently, the common belief was that the constraints from the
electron and neutron electric dipole moments are
strong~\cite{STRONGCP} and the new phases must be very small.  More
recent calculations performed in the framework of the minimal
supergravity model~\cite{OLIVE,NATH,BARTL} and non-minimal
models~\cite{KANE} indicated the possibility of cancellations between
contributions proportional to the phase of $\mu$ and those
proportional to the phase of $A$ and, therefore, of weaker limits on
the phases in a non-negligible range of parameter space.  The detailed
analysis~\cite{PRS} showed that the constraints on the phases
(particularly on the phase of $\mu$ and of the gaugino masses) are
generically strong $(\phi\leq 10^{-2})$ if all relevant supersymmetric
masses are light, say below ${\cal O}(300~\GeV)$.  However, the
constraints disappear or are substantially relaxed if just some of
those masses, e.g. slepton and sneutrino masses, are large, $m_E>
{\cal O} (1~\TeV)$.  Thus, the phases can be large even if some
masses, e.g. the chargino masses, are small.  In the parameter range
where the constraints are generically strong, there exist fine-tuned
regions where cancellations between different contributions to the EDM
can occur even for large phases.  They require not only $\mu-A$,
$\mu-M_{gaugino}$ or $M_1-M_2$ phase adjustments but also values of
soft mass parameters strongly correlated with the phases and among
themselves (especially for higher values of $\tan\beta$, as the
constraints on $\mu$ phase scale as $1/\tan\beta$).  Nevertheless,
since the notion of fine tuning is not precise, particularly from the
point of view of GUT models, it is not totally inconceivable that the
rationale for large cancellations exists in the large energy scale
physics~\cite{BHRLIK}.  Therefore all experimental bounds on the
supersymmetric parameters should include the possibility of
substantial phases allowing the possibility of large cancellations, to
claim full model independence.

We define the new flavor-conserving phases in the MSSM as:
\bea
e^{i\phi_{\mu}} = \frac{\mu}{|\mu|}
\hskip 15mm
e^{i\phi_i} = \frac{M_i}{|M_i|}
\hskip 15mm
e^{i\phi_{A_I}} = \frac{A_I}{|A_I|}
\hskip 15mm
e^{i\phi_H} = \frac{m_{12}^2}{|m_{12}^2|}
\label{eq:zpmphi}
\eea
Not all of those phases are physical (see~\cite{PRS} for a more
detailed discussion).  Physics observables depend only on the phases
of some parameter combinations.  Such combinations are:
\bea
M_i \mu (m^2_{12})^{\star} \hskip 2cm A_I \mu
(m^2_{12})^{\star} \hskip 2cm A_I^{\star} M_i
\label{eq:phasinv}  
\eea
Not all of them are independent: two of the phases can be rotated
away.  We follow the common choice and keep $m_{12}^2$ real in order
to have real tree level Higgs field VEV's and
$\tan\beta$\footnote{Loop corrections to the effective potential
induce phases in VEV's even if they were absent at the tree level.
Rotating them away reintroduces a phase into the $m_{12}^2$
parameter.}.  The second re-phasing may be used e.g. to make one of
the gaugino mass terms real - we choose it to be $M_2$.  With this
choice, chargino production cross section is sensitive only to the
$\mu$ parameter phase.  Neutralino production and chargino/neutralino
decay rates may depend additionally on the $M_1$ phase, but for the
purpose of this analysis we assume universal gaugino masses at the GUT
scale.  In such a case low-energy values of $M_1$ and $M_2$ are
connected by the relation $M_1 = 5/3\tan^2\theta_W M_2$ and the
identical phases of $M_1$ and $M_2$ can be simultaneously rotated
away.  In section~\ref{sec:zedm} we discuss the possible effects of
the departure from this assumption.

This paper is organized as follows.  In section~\ref{sec:xsec} we
present the general expressions for chargino and neutralino production
cross section for the case of complex couplings.  In
section~\ref{sec:scan} we define the range of scan over the MSSM
parameters we use and we discuss the possible effects of phases on
chargino masses, production and decay rates.  In
section~\ref{sec:leprev} we present the results of our scan, comparing
the expected chargino production and decay rates to the experimental
results obtained by DELPHI.  Several points where the introduction of
complex parameters could endanger the model-independence of the
experimental limits are found.  Then, in section~\ref{sec:zedm}, we
examine ways to restore the real-parameter exclusion limits by the use
of the EDM measurements as an additional constraint.  We finally
present our conclusions in section~\ref{sec:conclusions}. The
necessary conventions and Feynman rules are collected in
the~\ref{app:lagr}.

\section{Cross sections for the chargino and neutralino production}
\label{sec:xsec}

In this sections we list the formulae for the chargino and neutralino
production cross section in $e^+e^-$ collisions for the general case
of complex couplings.  For completeness, here and in
the~\ref{app:lagr} we give the most general expressions, including
also the possibility of flavor mixing of sneutrinos (we neglect only
the very small right-handed couplings of charginos to electrons,
$S_{RC}\sim{\cal O}\left(\frac{em_e\tan\beta}{M_W}\right)$).  However,
in our numerical analysis we consider only the simplified case,
neglecting possible inter-generational sneutrino mixing.

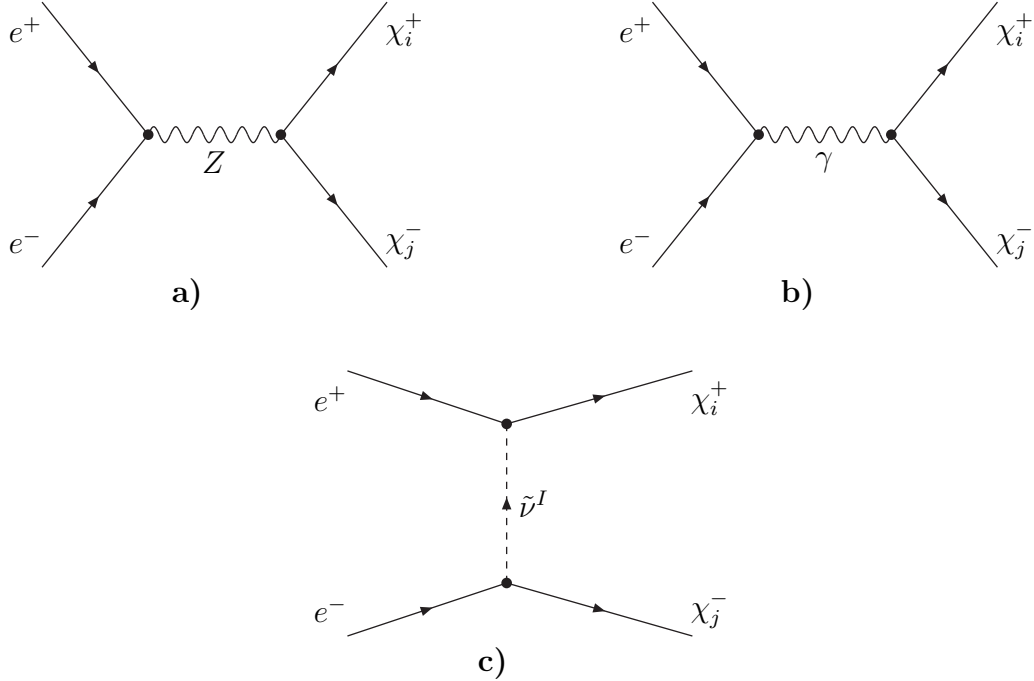
\begin{figure}[htbp]
\begin{center}
\begin{tabular}{lp{20mm}r}
\begin{picture}(150,120)(0,0)
\ArrowLine(20,110)(60,60)
\Text(20,100)[r]{\mbox{$e^+$}}
\ArrowLine(20,10)(60,60)
\Text(20,20)[r]{\mbox{$e^-$}}
\Vertex(60,60){2}
\Photon(60,60)(110,60){3}{6}
\Text(85,50)[c]{\mbox{$Z$}}
\Vertex(110,60){2}
\ArrowLine(110,60)(150,110)
\Text(150,100)[l]{\mbox{$\chi^+_i$}}
\ArrowLine(110,60)(150,10)
\Text(150,20)[l]{\mbox{$\chi^-_j$}}
\Text(75,0)[c]{\bf a)}
\end{picture}
&&
\begin{picture}(150,120)(0,0)
\ArrowLine(20,110)(60,60)
\Text(20,100)[r]{\mbox{$e^+$}}
\ArrowLine(20,10)(60,60)
\Text(20,20)[r]{\mbox{$e^-$}}
\Vertex(60,60){2}
\Photon(60,60)(110,60){3}{6}
\Text(85,50)[c]{\mbox{$\gamma$}}
\Vertex(110,60){2}
\ArrowLine(110,60)(150,110)
\Text(150,100)[l]{\mbox{$\chi^+_i$}}
\ArrowLine(110,60)(150,10)
\Text(150,20)[l]{\mbox{$\chi^-_j$}}
\Text(75,0)[c]{\bf b)}
\end{picture}
\end{tabular}
\vskip 5mm
\begin{picture}(150,120)(0,0)
\ArrowLine(20,110)(80,90)
\Text(20,100)[r]{\mbox{$e^+$}}
\ArrowLine(20,10)(80,30)
\Text(20,20)[r]{\mbox{$e^-$}}
\Vertex(80,90){2}
\DashArrowLine(80,30)(80,90){3}
\Text(85,60)[l]{\mbox{$\tilde{\nu}^I$}}
\Vertex(80,30){2}
\ArrowLine(80,90)(150,110)
\Text(150,100)[l]{\mbox{$\chi^+_i$}}
\ArrowLine(80,30)(150,10)
\Text(150,20)[l]{\mbox{$\chi^-_j$}}
\Text(75,0)[c]{\bf c)}
\end{picture}
\vskip 5mm
\caption{Diagrams contributing to chargino production in $e^+e^-$
collisions.}
\label{fig:charprod}
\end{center}
\end{figure}

The chargino production amplitude is given by the three diagrams shown
in fig.~\ref{fig:charprod}.  Using the notation defined
in~\ref{app:lagr}, the differential cross section in the CMS frame for
the process $e^+e^-\ra \chi^+_i\chi^-_j$ can be written as:
\bea
{d\sigma(e^+e^-\ra \chi^+_i \chi^-_j)\over d\Omega} =
{\lambda(s,m_i^2,m_j^2)\over 64\pi^2 s^2} \left(M_{aa} + M_{bb} + M_{cc}
+ M_{ab} + M_{ac} + M_{bc}\right)
\label{eq:zcc_diff}
\eea
where $m_i\equiv m_{\chi^+_i}$, $m_j\equiv m_{\chi^+_j}$,
$\lambda(x,y,z) = (x^2 + y^2 + z^2 -2xy -2xz -2yz)^{1/2}$ and $M_{xy}$
respond to contributions of respective diagrams in
fig.~\ref{fig:charprod} and their interference:
\bea
M_{aa}&=& {e^2 \over 4 |D_Z(s)|^2}\left[(a_e^2 + b_e^2)
\left((|V_{LC}^{ij}|^2+ |V_{RC}^{ij}|^2)\left(s^2 - (m_i^2
- m_j^2)^2 +\lambda^2\cos^2\theta\right)\right.\right.\nonumber\\
&+&\left.\left.8s m_i m_j\mathrm{Re}
(V_{LC}^{ij}V_{RC}^{ij\star})\right) - 2(a_e^2 -  b_e^2)
(|V_{LC}^{ij}|^2 - |V_{RC}^{ij}|^2) s\lambda\cos\theta \right]
\label{eq:zcc_diffz}\\
M_{bb}&=& {e^4 \over s^2} \left(s^2 + 4 s m_i^2 
+\lambda^2\cos^2\theta\right)\delta_{ij}
\label{eq:zcc_diffph}\\
M_{cc}&=& {e^4\over 16 s_W^4 |D_{\tilde{\nu}}(t)|^2}
|Z_+^{1i}Z_+^{1j}|^2
\left[(s - \lambda\cos\theta)^2 - (m_i^2 - m_j^2)^2\right]
\label{eq:zcc_diffs}\\
M_{ab}&=& {e^3(s-M_Z^2)\over 2 s |D_Z(s)|^2}
\left((a_e+b_e)(V_{LC}^{ii} + V_{RC}^{ii}) \left(s^2 + 4s m_i^2 
 + \lambda^2\cos^2\theta\right) \right.\nonumber\\
&-& \left.2(a_e-b_e) (V_{LC}^{ii} - V_{RC}^{ii}) s \lambda
\cos\theta\right) \delta_{ij}
\label{eq:zcc_diffzph}\\
M_{ac}&=& {e^3 a_e\over 4s_W^2} \mathrm{Re} \left( {V_{LC}^{ij}\left((s -
\lambda\cos\theta)^2 - (m_i^2 - m_j^2)^2\right) + 4 V_{RC}^{ij} s m_i
m_j \over D_Z^{\star}(s)D_{\tilde{\nu}}(t)}Z_+^{1i} Z_+^{1j\star}\right)
\label{eq:zcc_diffzs}\\
M_{bc}&=& {e^4\over 4 s_W^2 s D_{\tilde{\nu}}(t)} |Z_+^{1i}|^2 
\left((s - \lambda\cos\theta)^2  + 4 s m_i^2\right)\delta_{ij}
\label{eq:zcc_diffphs}
\eea
By $a_e, b_e$ we denoted the left and right part of the $Z\bar e e$
coupling, $ie(a_e P_L + b_e P_R)$ (so that $a_e = (2s_W^2 -
1)/2s_Wc_W$, $b_e = s_W/c_W$), $D_Z(s) = s - M_Z^2 + iM_Z\Gamma_Z$, $t
= \frac{1}{2}(m_i^2 + m_j^2 - s + \lambda\cos\theta)$ and
$D_{\tilde{\nu}}(t)$ is ``flavor averaged'' $t$-channel sneutrino
propagator (for vanishing sneutrino mixing it reduces simply to
electron sneutrino propagator):
\bea
{1\over D_{\tilde{\nu}}(t)} = \sum_{I=1}^3 {|Z_{\tilde{\nu}}^{1I}|^2\over t -
m_{\tilde{\nu}^I}^2}
\eea

The total cross section for the chargino pair production reads as:
\bea
\sigma(e^+e^-\ra \chi^+_i \chi^-_j) = {\lambda(s,m_i^2,m_j^2)\over
16\pi s^2} \left({\cal M}_{aa} + {\cal M}_{bb} + {\cal M}_{cc} + {\cal
M}_{ab} + {\cal M}_{ac} + {\cal M}_{bc}\right)
\label{eq:zcc}
\eea
where ${\cal M}_{xy}$ are:
\bea
{\cal M}_{aa}&=& {e^2(a_e^2 + b_e^2)\over 4 |D_Z(s)|^2} \left[
(|V_{LC}^{ij}|^2+|V_{RC}^{ij}|^2)\left(s^2 - (m_i^2 - m_j^2)^2
+\f{\lambda^2}{3}\right) \right.\\
&+&\left.8s m_i m_j\mathrm{Re} (V_{LC}^{ij}V_{RC}^{ij\star})\right]
\label{eq:zcc_z}\\
{\cal M}_{bb}&=& {4e^4 \over 3s } \left(s + 2 m_i^2\right)\delta_{ij}
\label{eq:zcc_ph}\\
{\cal M}_{cc}&=& {e^4 \over 4s_W^4} |Z_+^{1i}Z_+^{1j}|^2
\sum_{I,J=1}^3 |Z_{\tilde{\nu}}^{1I} Z_{\tilde{\nu}}^{1J}|^2
\left[1 \right.\nonumber\\
&+& \left.2 {(m_{\tilde{\nu}^I}^2 - m_i^2)(m_{\tilde{\nu}^I}^2 - m_j^2)
L(m_{\tilde{\nu}^I}^2) 
- (m_{\tilde{\nu}^J}^2 - m_i^2)(m_{\tilde{\nu}^J}^2 - m_j^2)
L(m_{\tilde{\nu}^J}^2) \over \lambda(m_{\tilde{\nu}^I}^2 -
m_{\tilde{\nu}^J}^2)}\right]
\label{eq:zcc_s}\\
{\cal M}_{ab}&=& {2e^3(a_e+b_e)(s-M_Z^2)\over 3 |D_Z(s)|^2}
(V_{LC}^{ii} + V_{RC}^{ii}) (s + 2 m_i^2) \delta_{ij}
\label{eq:zcc_zph}\\
{\cal M}_{ac}&=& {e^3a_e\over 2s_W^2}\sum_{I=1}^3
|Z_{\tilde{\nu}}^{1I}|^2
\mathrm{Re}\left[ {Z_+^{1i}Z_+^{1j\star} \over D_Z^{\star}(s)}\left(
V_{LC}^{ij}\left(2 m_{\tilde{\nu}^I}^2 - m_i^2 - m_j^2
-s\right.\right.\right.\nonumber\\
& +&\left.\left.\left.{2 ((m_{\tilde{\nu}^I}^2 -
m_i^2)(m_{\tilde{\nu}^I}^2 - m_j^2))\over \lambda}
L(m_{\tilde{\nu}^I}^2)\right) + V_{RC}^{ij} {2 s m_i m_j\over \lambda}
L(m_{\tilde{\nu}^I}^2) \right)\right]
\label{eq:zcc_zs}\\
{\cal M}_{bc}&=& {e^4\over s_W^2 s}\delta_{ij}|Z_+^{1i}|^2\sum_{I=1}^3 
|Z_{\tilde{\nu}}^{1I}|^2 \left[m_{\tilde{\nu}^I}^2 - m_i^2  -\f{1}{2}s
 + {(m_{\tilde{\nu}^I}^2 - m_i^2)^2 + s m_i^2\over \lambda}
L(m_{\tilde{\nu}^I}^2) \right]
\label{eq:zcc_phs}
\eea
where we defined function $L$ as
\bea
L(m_{\tilde{\nu}^I}^2) = \log\left({2 m_{\tilde{\nu}^I}^2 - m_i^2 -
m_j^2 + s -\lambda\over 2 m_{\tilde{\nu}^I}^2 - m_i^2 - m_j^2 + s
+\lambda }\right)
\eea

For the purpose of this paper we use only the neutralino production
cross section at the $Z^0$ peak, in order to calculate the invisible
$Z^0$ decay width.  Thus, it is sufficient to include only the
$s$-channel diagram in the expression for the neutralino production
amplitude.  In this approximation, the differential cross section in
the CMS frame for the process $e^+e^-\ra \chi^0_i\chi^0_j$ has a
simple form:
\bea
{d\sigma(e^+e^-\ra \chi^0_i \chi^0_j)\over d\Omega}& =& {(2 -
\delta_{ij})e^2 (a_e^2 + b_e^2) \lambda(s,m_i^2,m_j^2) \over 
64\pi^2 s^2|D_Z(s)|^2} \left[ |V_{N}^{ij}|^2
\left(s^2 - (m_i^2 - m_j^2)^2 +\lambda^2\cos^2\theta\right) \right.\nonumber\\
&-&\left.4s m_i m_j\mathrm{Re} (V_{N}^{ij})^2 \right]
\label{eq:znn_diff}
\eea
where now $m_i\equiv m_{\chi^0_i}$, $m_j\equiv m_{\chi^0_j}$.

The total cross section for the process $e^+e^-\ra \chi^0_i\chi^0_j$
reads as:
\bea
\sigma(e^+e^-\ra \chi^0_i \chi^0_j) &=& {(2 - \delta_{ij})
e^2 (a_e^2 + b_e^2) \lambda(s,m_i^2,m_j^2)\over 16\pi s^2|D_Z(s)|^2} 
\left[|V_{N}^{ij}|^2 \left(s^2 - (m_i^2 - m_j^2)^2 +{\lambda^2\over
3}\right)\right.\nonumber\\
& -&\left.4s m_i m_j\mathrm{Re} (V_{N}^{ij})^2 \right]
\label{eq:znn}
\eea

\section{Features of the chargino searches for the complex MSSM parameters}
\label{sec:scan}

In order to check the effects of introduction of complex couplings on
LEP limits, we performed a scan over the parameters present in the
gaugino mass matrices and the chargino couplings.  These parameters
are:
\begin{itemize}
\item $|\mu|$, modulus of the Higgs mixing parameter.  We assumed it
  running from 5 to 500 GeV, with a step of 5 GeV.
\item $M_2$, the $SU(2)$ gaugino mass.  We scan over it in the same
  range as for the parameter $|\mu|$: 5 to 500 GeV and the same step.
  We also assume that the GUT relation between the gaugino masses
  holds: $M_1 = \frac{5}{3}\tan\theta_W^2 M_2$.  In this case, the
  common gaugino mass parameter phase can be rotated away and both
  $M_1,M_2$ can be chosen to be real.
\item $m_{\tilde{\nu}_e}$, the mass of the electron sneutrino,
  contributing to the $t$-channel diagram in the chargino production
  cross section.  We consider the cases for which the sneutrino mass
  is 45, 50, 60, 70, 80, 90, 100, 110, 200 and 300 GeV.
\item $\tan\beta$ parameter.  We consider three values: two small ones
  $\tan\beta=1,1.5$ and one large $\tan\beta=35$.
\item Finally, we perform the scan over $\phi_\mu$, the phase of the
  $\mu$ parameter, from 0 to $\pi$ with a step of $\pi/18$ (variation
  of $\phi_\mu$ in the extended range $0-2\pi$ leads to identical
  results).
\item The right selectron mass was fixed to be at high value (300
  GeV).  It only comes in the chargino decay to neutralino plus
  leptons, and might alter some of the branching fractions, but since
  we already scan over the sneutrino and therefore the left selectron
  mass, the possibility of a light charged slepton is taken into
  account.  It is also worth noting that the assumption of heavy right
  selectron is a conservative one for what concerns the EDM
  constraints on $\mu$ phase.
\end{itemize}
The output of our scan are the physical gaugino masses, the chargino
cross sections and branching ratios for a center mass energy of 189
GeV as well as the neutralino and chargino contributions to the $Z^0$
width.  We computed these using the SUSYGEN~3 program~\cite{SUSYGEN},
whose independent calculation of the chargino and neutralino
production cross sections has been checked against the formulae in
section~\ref{sec:xsec}.

\subsection{Effects of complex parameters on physical masses}
\label{subsec:masses}

\begin{figure}[htbp]
\begin{center}
\includegraphics[width=0.48\linewidth]{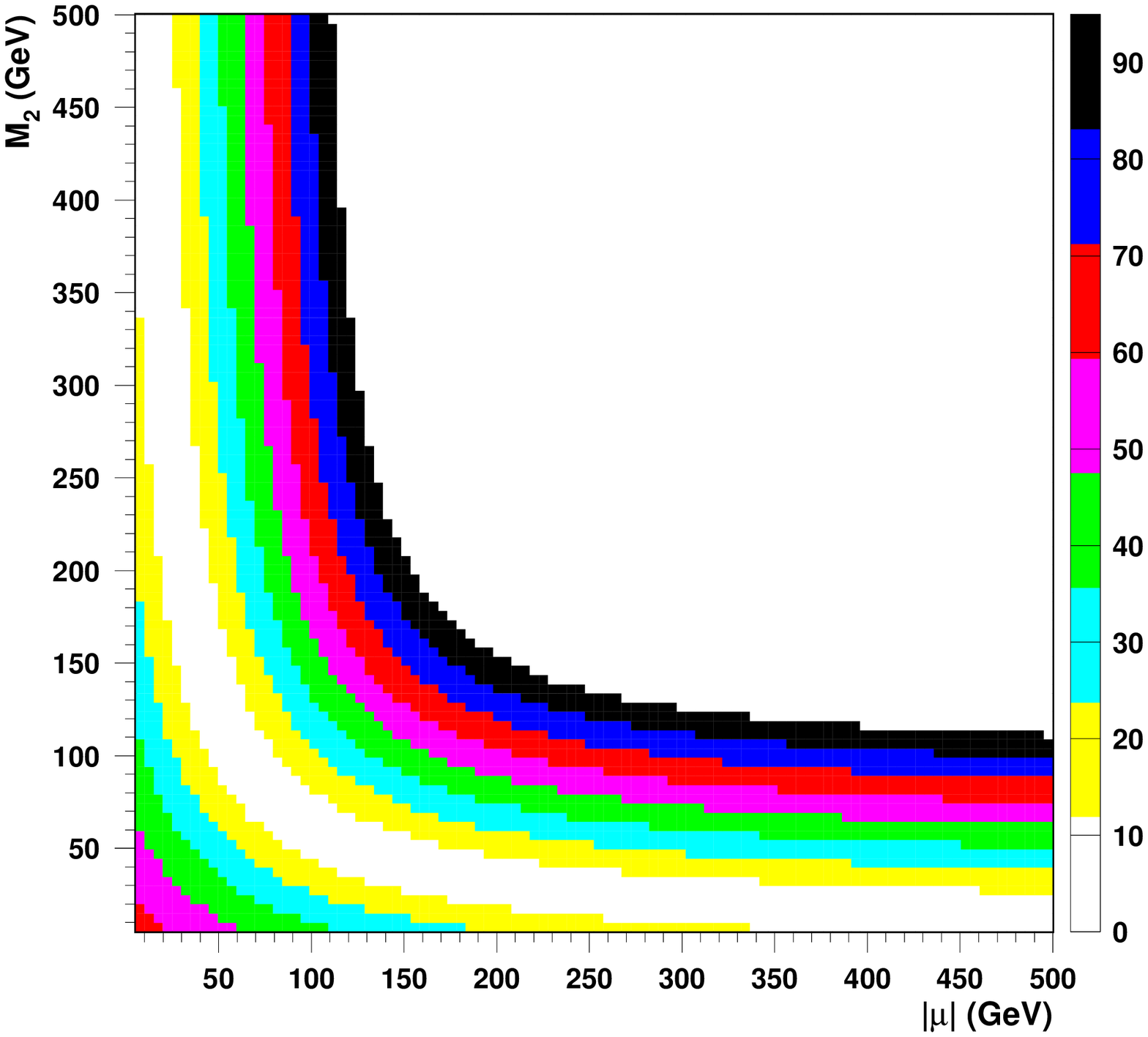}
\includegraphics[width=0.48\linewidth]{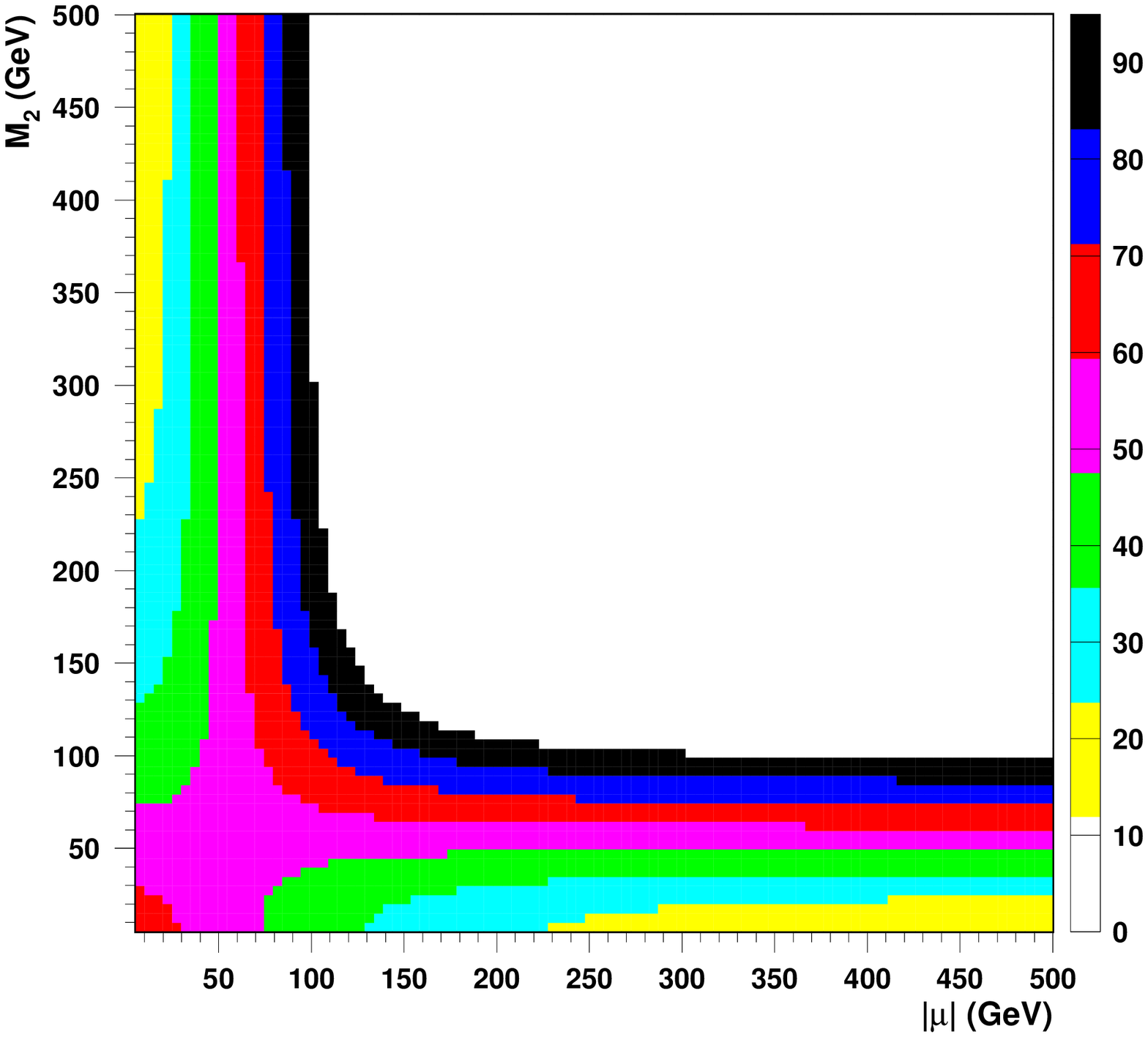}
\includegraphics[width=0.48\linewidth]{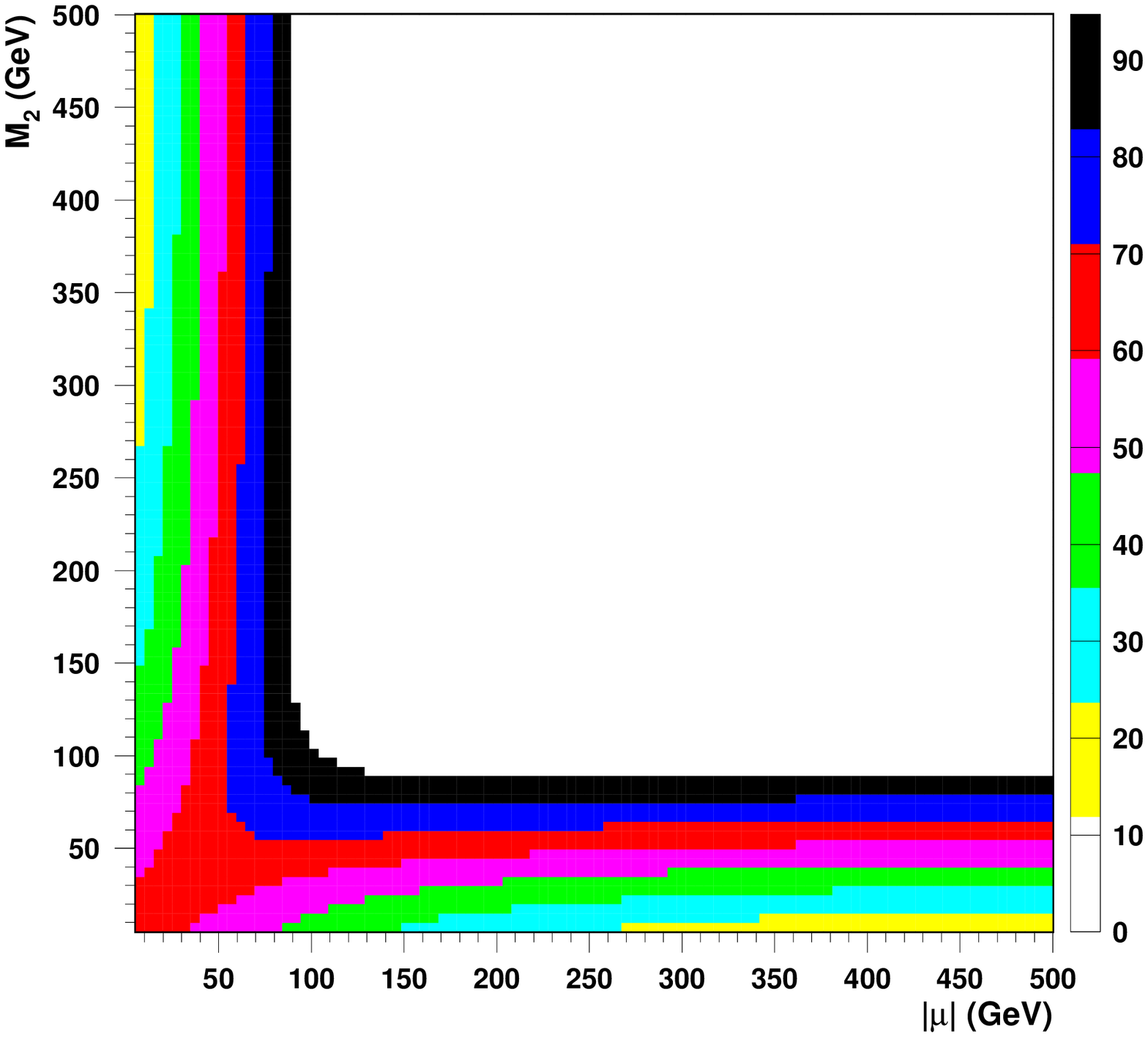}
\includegraphics[width=0.48\linewidth]{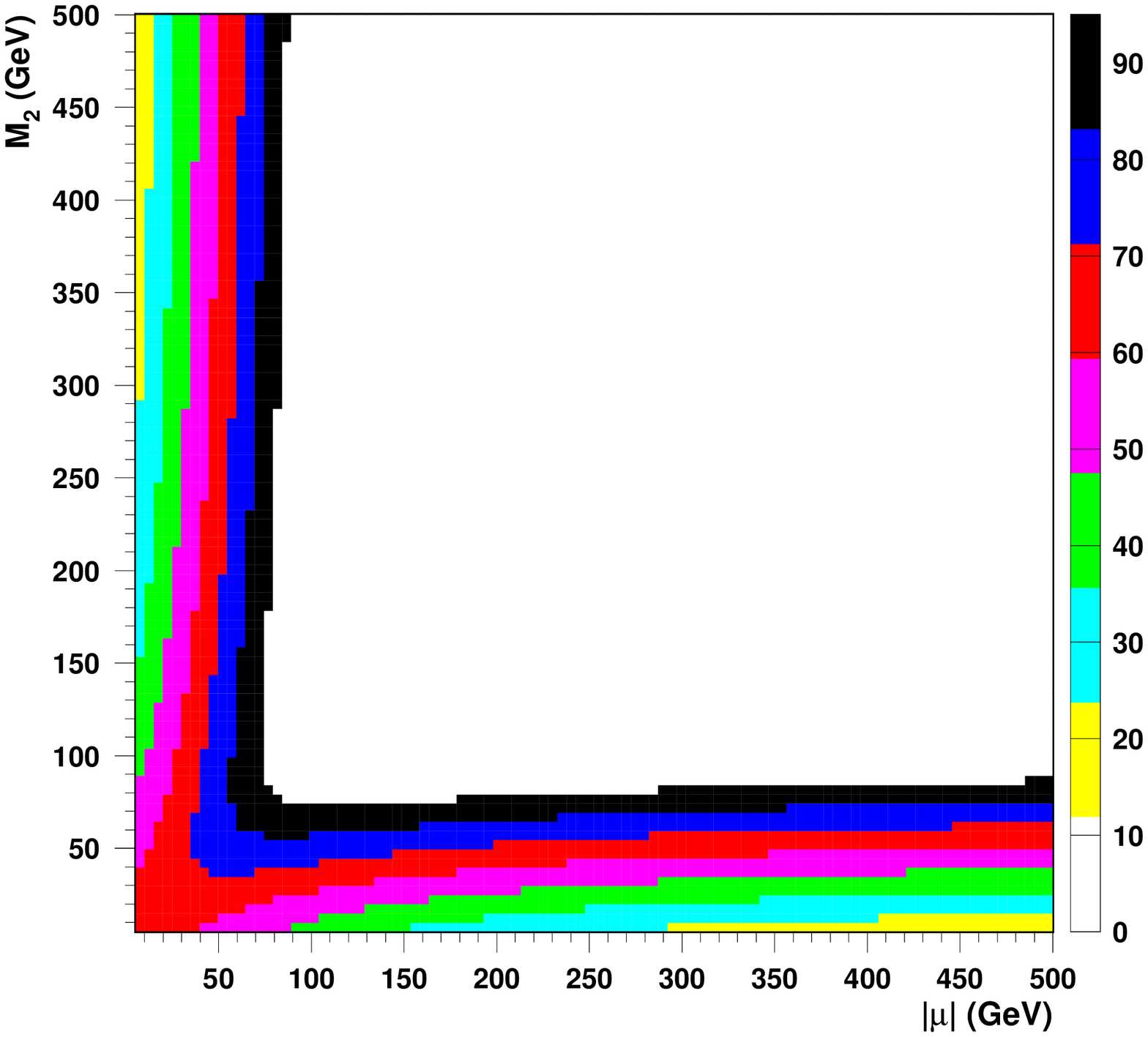}
\caption{Contours of the constant chargino mass on the $(M_2,|\mu|)$
plane for $\tan\beta=1.5$ and $\phi_{\mu}=0,\pi/2,3\pi/4,\pi$.}
\label{fig:cmassdist}
\end{center}
\end{figure}

\begin{figure}[htbp]
\begin{center}
\includegraphics[width=0.48\linewidth]{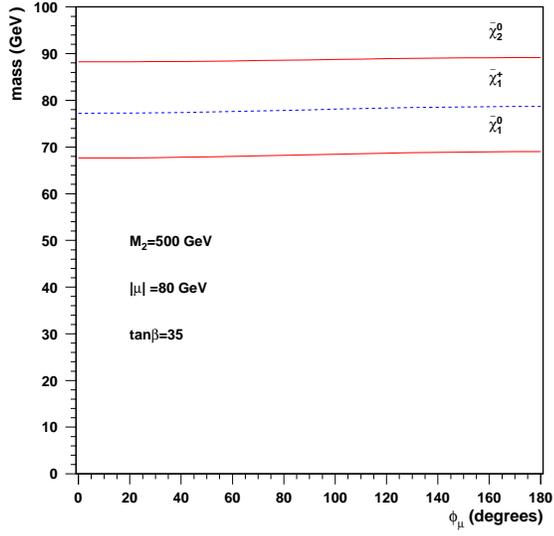}
\includegraphics[width=0.48\linewidth]{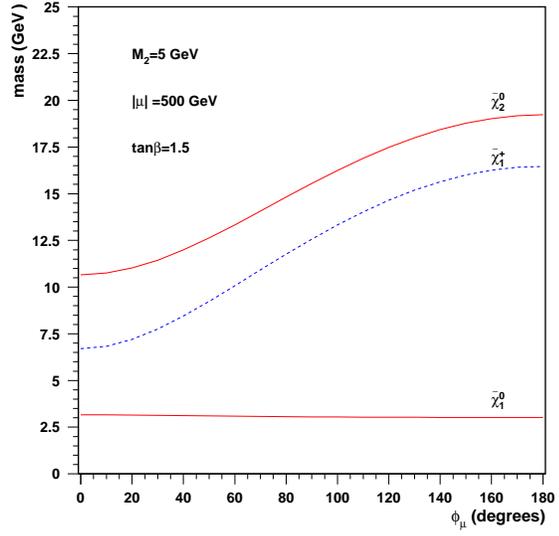}
\includegraphics[width=0.48\linewidth]{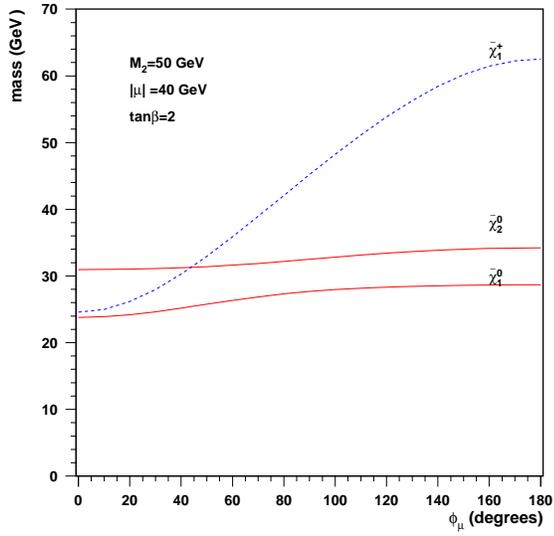}
\includegraphics[width=0.48\linewidth]{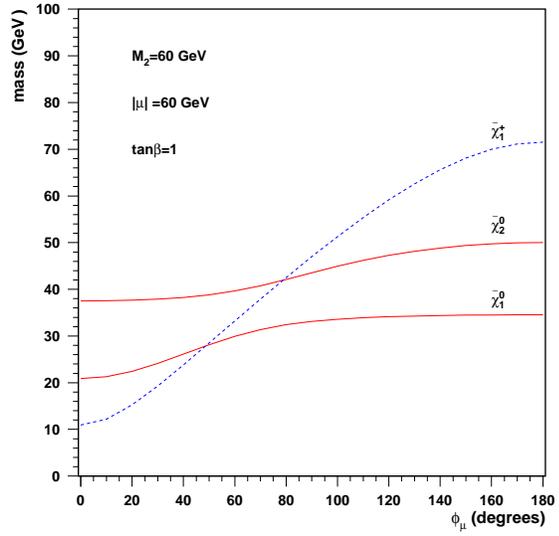}
\caption{Chargino and neutralino mass evolution as a function of
$\phi_{\mu}$.}
\label{fig:masses}
\end{center}
\end{figure}

Figure~\ref{fig:cmassdist} shows the chargino mass distributions in
the ($M_2,|\mu|$) plane for chosen values of the $\mu$ phase and a low
value of $\tan\beta$.  The first feature worth noting is that the
chargino mass never decreases with the $\mu$ phase increasing from $0$
(real positive $\mu$) to $\pi$ (real negative $\mu$) for any value of
the remaining MSSM parameters.  This is even more obvious from
figure~\ref{fig:masses}, where the chargino and neutralino mass
dependence on $\phi_{\mu}$ is shown for selected values of the
remaining MSSM parameters.  Another feature that one can notice in
figure~\ref{fig:cmassdist} is the presence of the well known region of
very low chargino masses for small $\phi_{\mu}$ values and the
emergence of a high chargino mass corridor for $M_2\sim|\mu|\tan\beta$
and increasing $\phi_{\mu}$.

\begin{figure}[htbp]
\begin{center}
\includegraphics[width=0.48\linewidth]{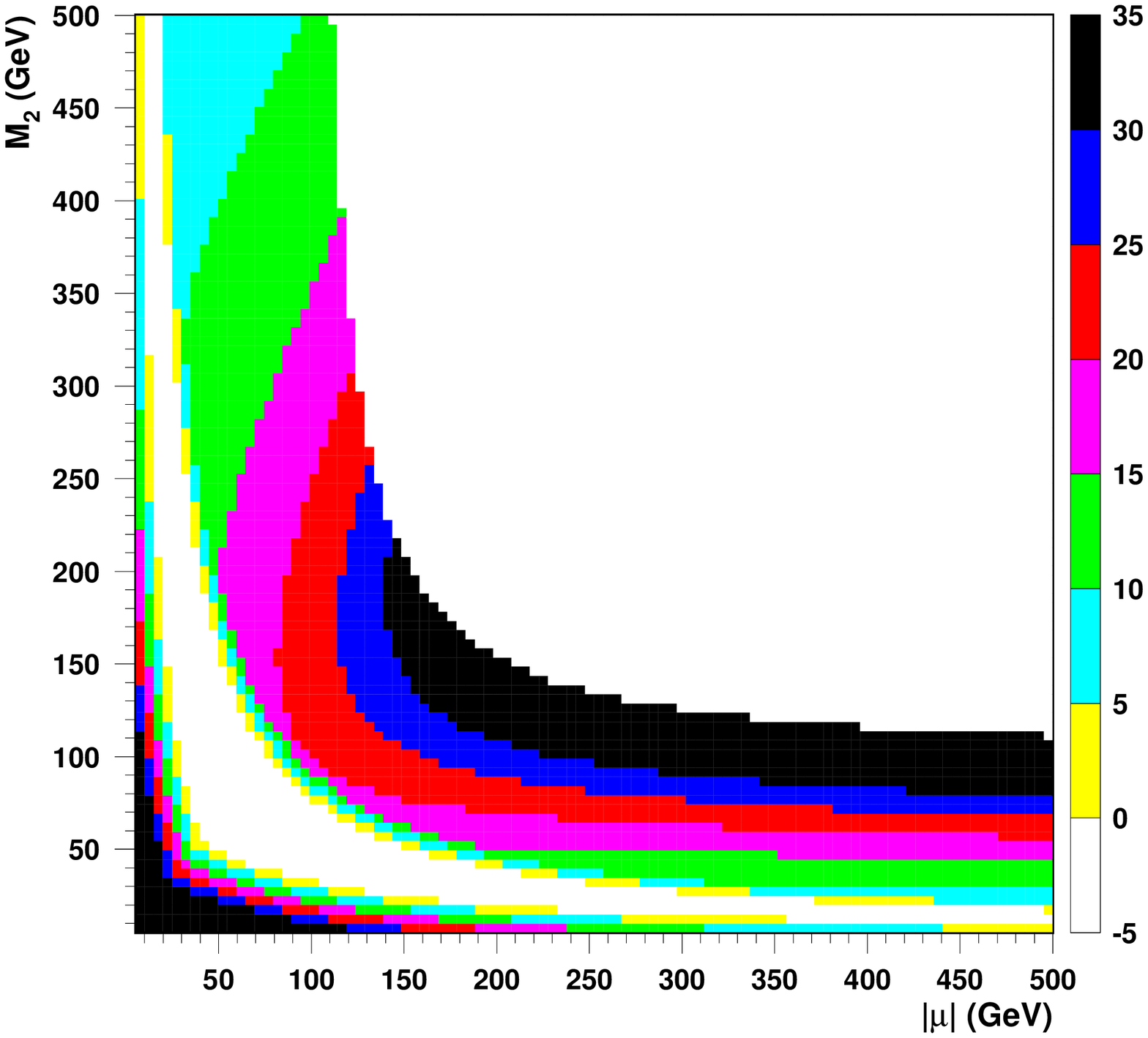}
\includegraphics[width=0.48\linewidth]{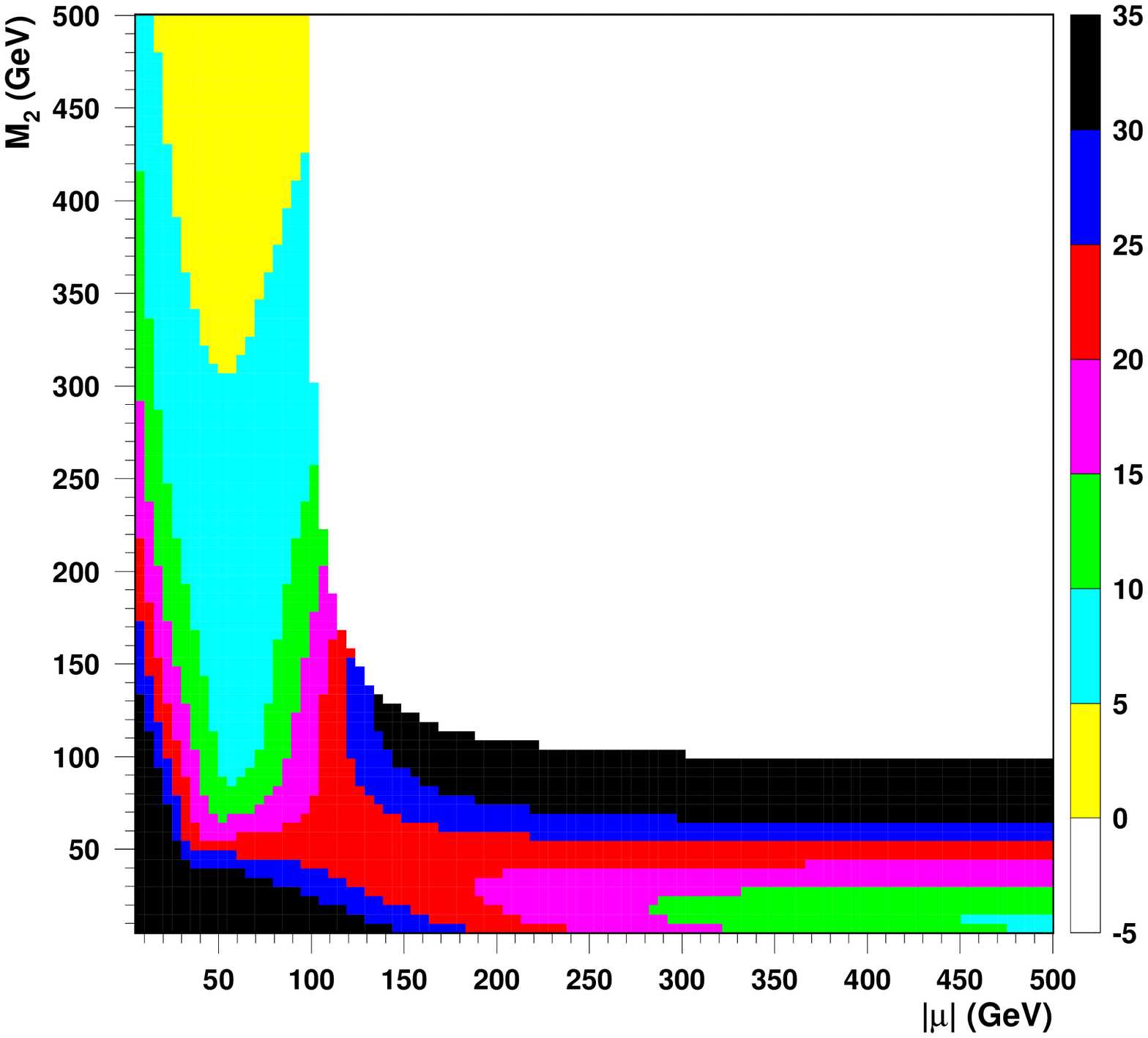}
\includegraphics[width=0.48\linewidth]{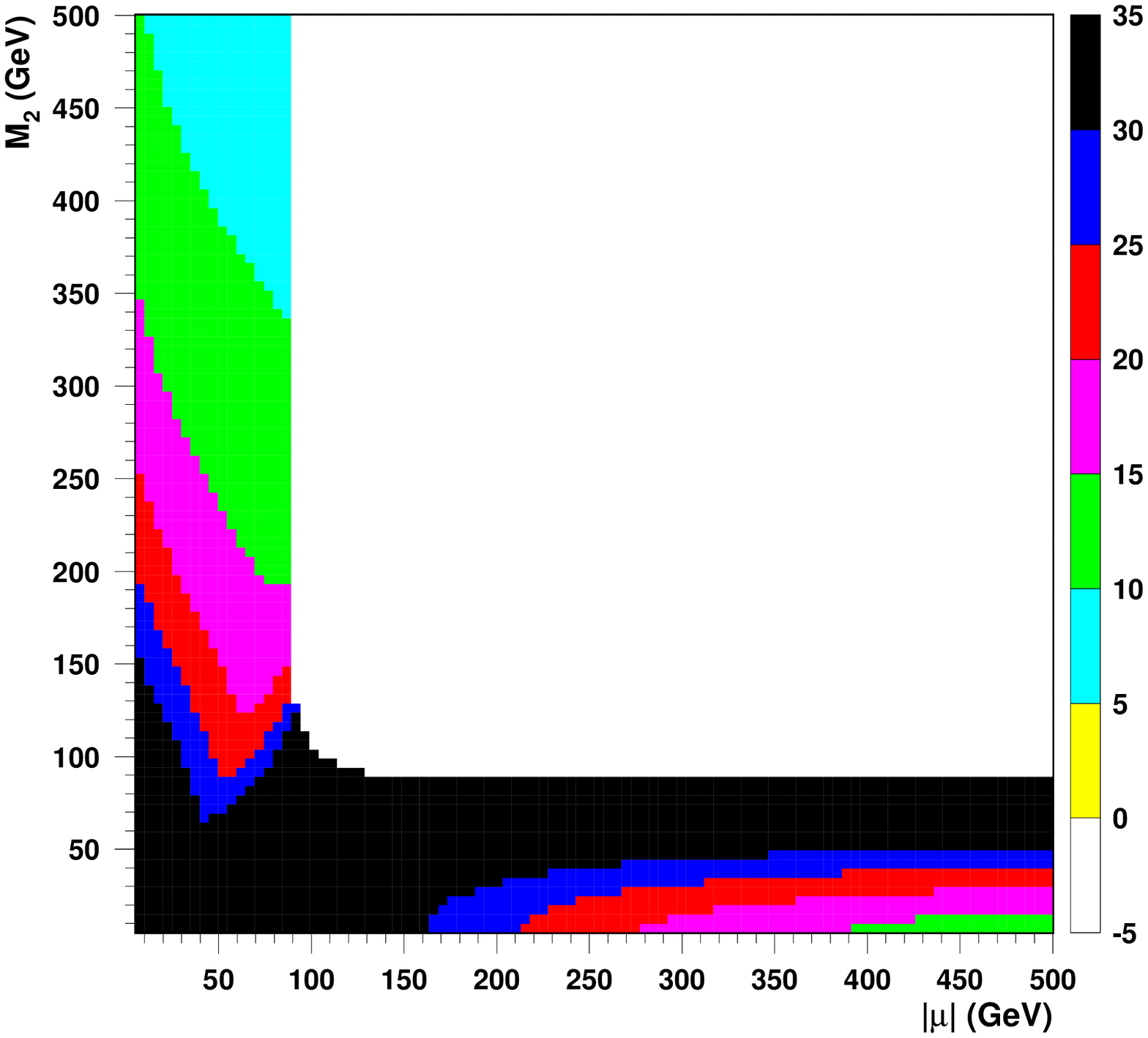}
\includegraphics[width=0.48\linewidth]{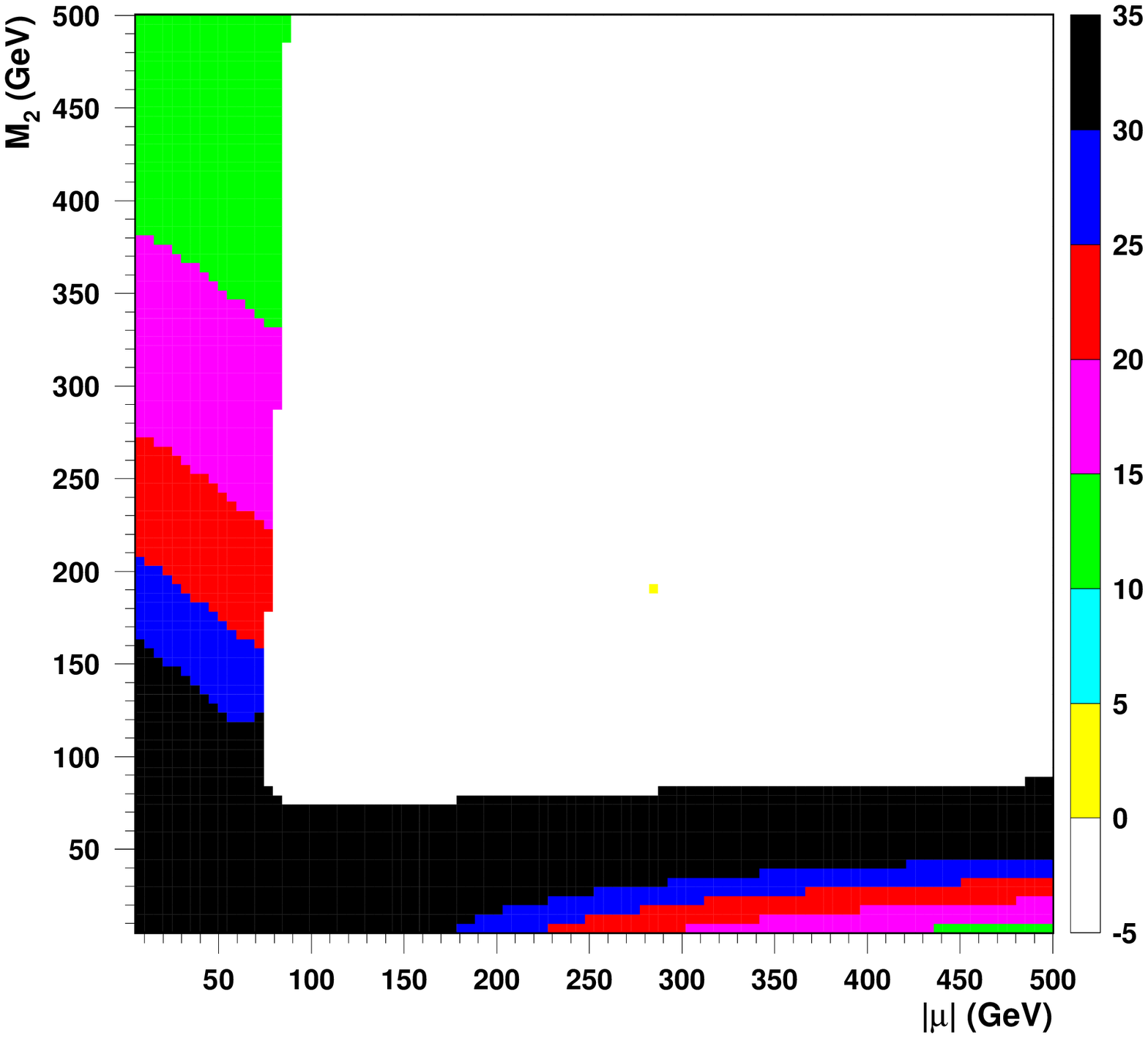}
\caption{Contours of the constant chargino-lightest neutralino mass
difference on the $(M_2,|\mu|)$ plane for $\tan\beta=1.5$ and
$\phi_{\mu}=0,\pi/2,3\pi/4,\pi$.}
\label{fig:diffdist}
\end{center}
\end{figure}

In figure~\ref{fig:diffdist} we also plotted contour lines of the
constant chargino-lightest neutralino mass splitting.  One can see
that for small $\phi_{\mu}$ there is an unphysical region where the
chargino becomes lighter than the neutralino.  One can also see the
striking feature that for pure imaginary $\mu$ the region of
chargino-neutralino degeneracy at high $M_2$ and low $|\mu|$ is
greatly enhanced, endangering the efficiency of the experimental
searches.

In general, one can observe the following types of relations between
the chargino and neutralino masses as a function of $\phi_{\mu}$:
\begin{itemize}
\item The chargino/neutralino masses depend weakly on $\phi_{\mu}$\ 
  for high $\tan\beta$, since then some of the off-diagonal elements
  in the mass matrices are very small, depressing the effects of $\mu$
  phase on physical masses (left upper plot in fig.~\ref{fig:masses}).
\item The chargino mass shows a dependence on $\phi_{\mu}$ whereas the
  $\chi^0_1$ mass evolves very slowly.  The chargino-neutralino mass
  difference is always positive and its largest value is obtained for
  real negative $\mu$ values (right upper plot in
  fig.~\ref{fig:masses}).
\item The same configuration can have the extreme case where for real
  positive $\mu$ ($\phi_\mu=0$) one has chargino almost degenerate
  with the lightest neutralino, whereas for real negative $\mu$, one
  has a large mass difference and even cascade decays through
  $\chi^0_2$ (fig.~\ref{fig:masses}, left lower plot).
\item Finally, as it has been already shown in the previous
  two-dimensional figures, and it is a well known fact since some
  time, one has a region where the chargino becomes lighter than the
  neutralino for low $M_2$ values, real positive $\mu$ and small
  $\tan\beta$.  In the right lower plot of fig.~\ref{fig:masses} one
  can see how the masses evolve with $\phi_{\mu}$ restoring the normal
  hierarchy between them.
\end{itemize}

\begin{figure}[htbp]
\begin{center}
\includegraphics[width=0.6\linewidth]{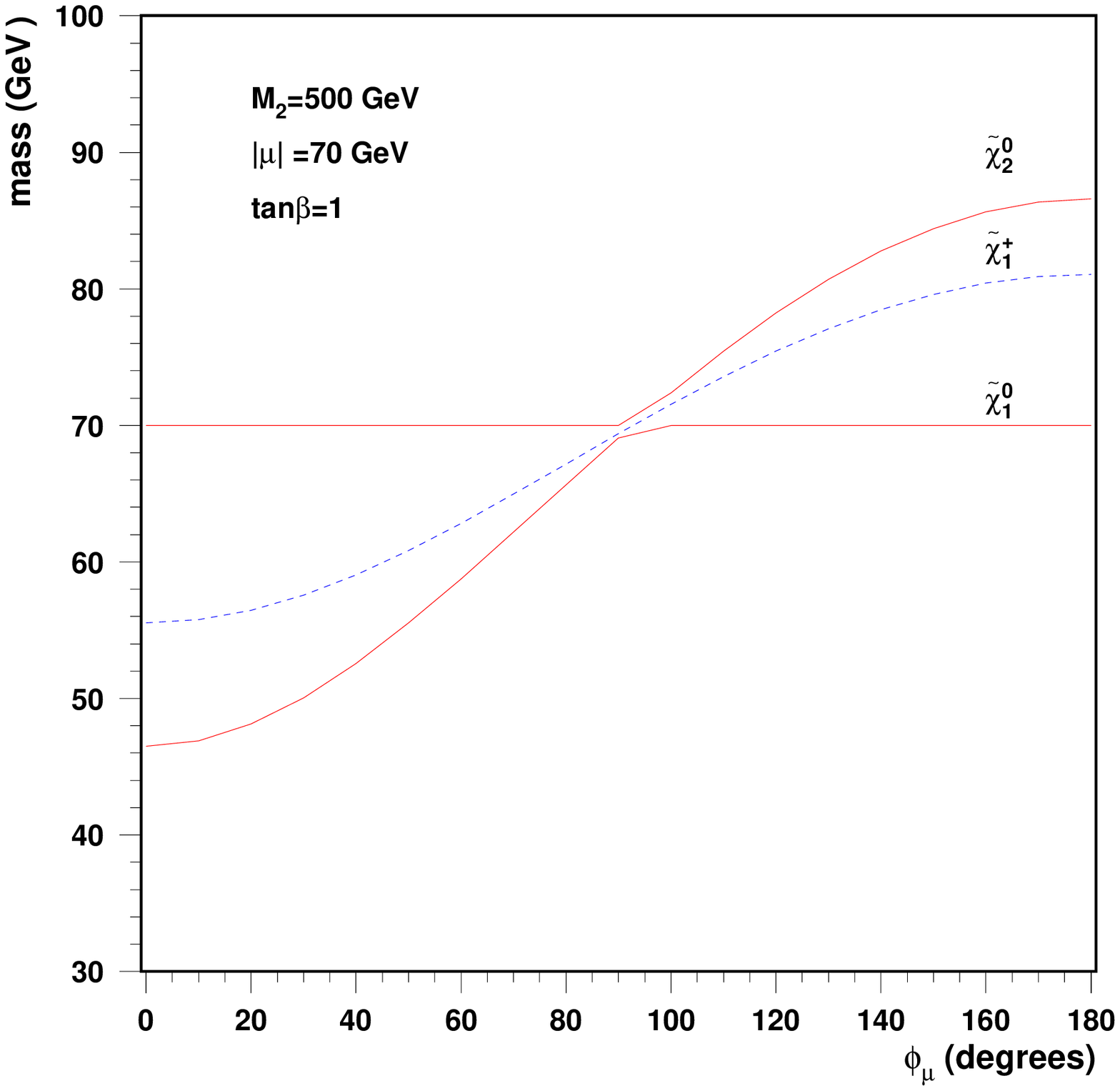}
\caption{Example of extreme chargino/neutralino mass degeneracy for
pure imaginary $\mu$.}
\label{fig:massdeg}
\includegraphics[width=0.6\linewidth]{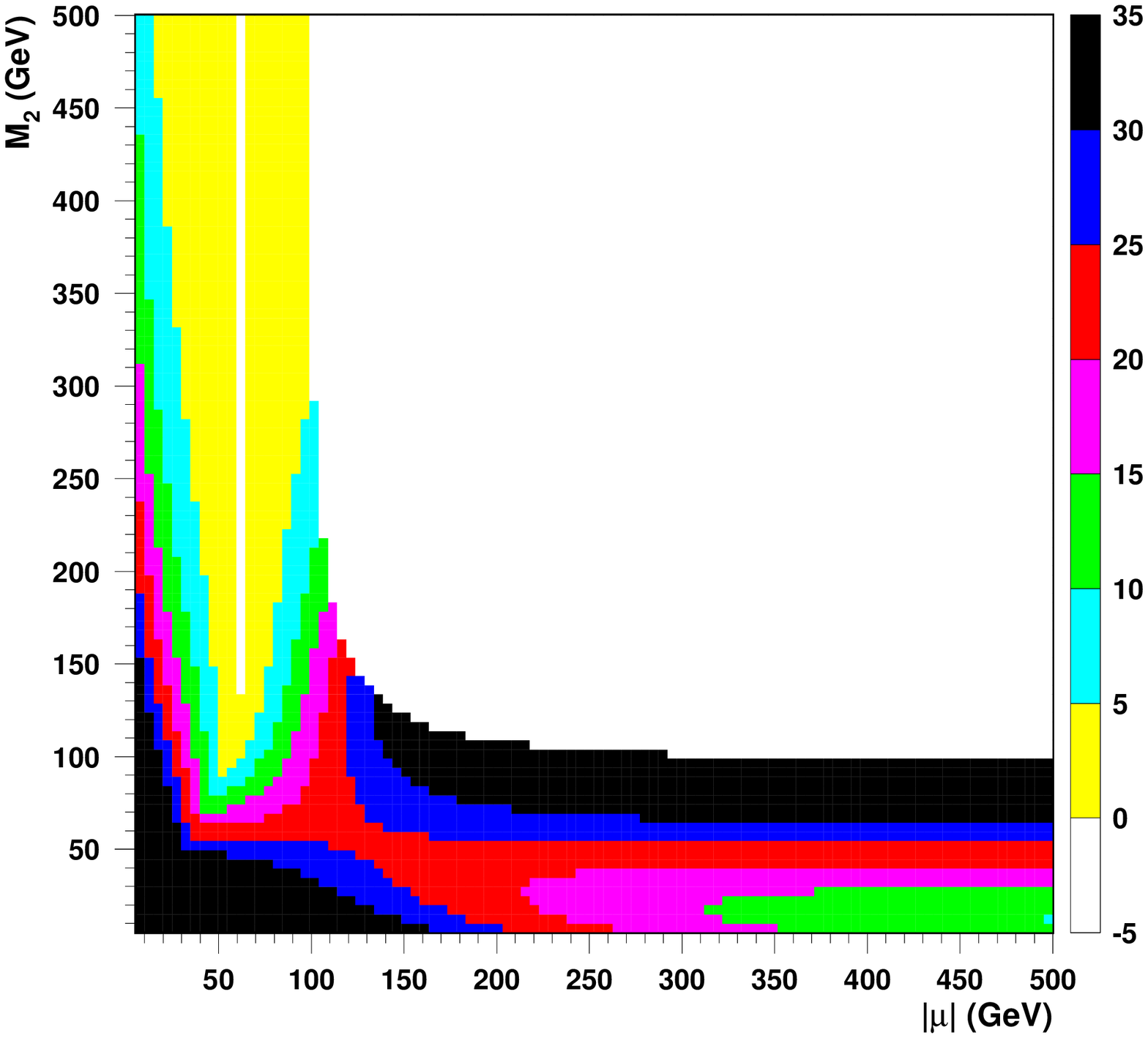}
\caption{Contours of the constant chargino-lightest neutralino mass
difference on the $(M_2,|\mu|)$ plane for $\tan\beta=1$ and
$\phi_{\mu}=\pi/2$.}
\label{fig:massdeg1}
\end{center}
\end{figure}

We systematically looked for areas, where the limits obtained by the
LEP experimental searches, assuming real $\mu$, could be endangered by
the introduction of the $\mu$ phase.  We thus searched for regions of
degeneracy between the chargino and lightest neutralino, located far
from the two extreme phase values ($0$ and $\pi$, giving real positive
and real negative $\mu$).  Needless to point out that degeneracies
like that make the experimental detection very difficult, if not
impossible, since the latter depends critically on the size of the
visible mass difference.  As it was shown generically in
figure~\ref{fig:diffdist} and more specifically in
figure~\ref{fig:massdeg}, one can find parameter values for which
neutralino and chargino masses are very closely degenerate for
$\phi_\mu=\pi/2$, i.e. for pure imaginary $\mu$.
Figure~\ref{fig:massdeg1} shows that this feature is characteristic of
a whole region around $|\mu|=70$GeV, for $\tan\beta=1$.

Such a situation occurs for $\tan\beta$ close to $1$ (a value which is
however already strongly disfavored by the direct SUSY Higgs particle
searches) and $M_2/|\mu|\gg 1$, i.e. when the lighter chargino and the
two lighter neutralinos are almost pure Higgsinos.  In such a case one
may estimate analytically that for real $\mu$ chargino-neutralino mass
difference is:
\bea
m^2_{\chi_1^+} - m^2_{\chi_1^0} \approx  \left\{
\begin{array}{cp{1cm}l}
{2m_Z^2 s_W^2 |\mu| \over M_1} & & \phi_{\mu}=0\\
&&\\ 
{2 m_W^2 |\mu| \over M_2 }& & \phi_{\mu}=\pi\\
\end{array}\right.
\label{eq:realdiff}
\eea
whereas for the pure imaginary $\mu$ the analogous formulae can be
written down as:
\bea
m^2_{\chi_1^+} - m^2_{\chi_1^0} &\approx& { |\mu|^2m_Z^2 s_W^2
\over M_1^2 - |\mu|^2} \approx {2 |\mu|^2 m_Z^2 s_W^2 \over M_1^2}
\label{eq:imagdiff}
\eea 
The mass difference for the $\phi_{\mu}=\pi/2$ is suppressed by one
more power of the $|\mu|/M_1$ ratio than for the real $\mu$, thus
decreases much faster for fixed $|\mu|$ and heavy gauginos.

\subsection{Effects of complex parameters on cross sections and
branching fractions}
\label{subsec:xcross}

\begin{figure}[htbp]
\begin{center}
\includegraphics[width=0.48\linewidth]{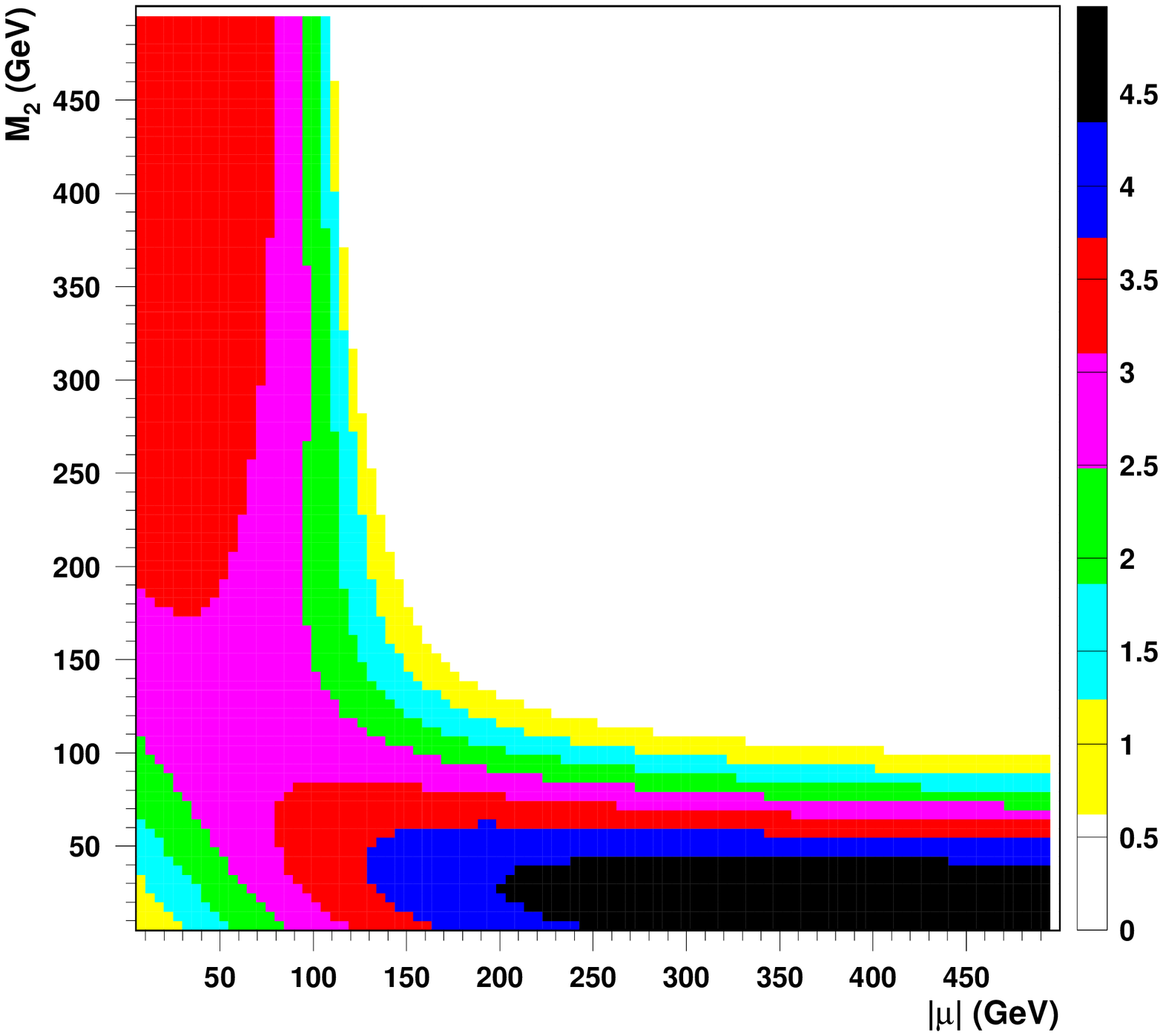}
\includegraphics[width=0.48\linewidth]{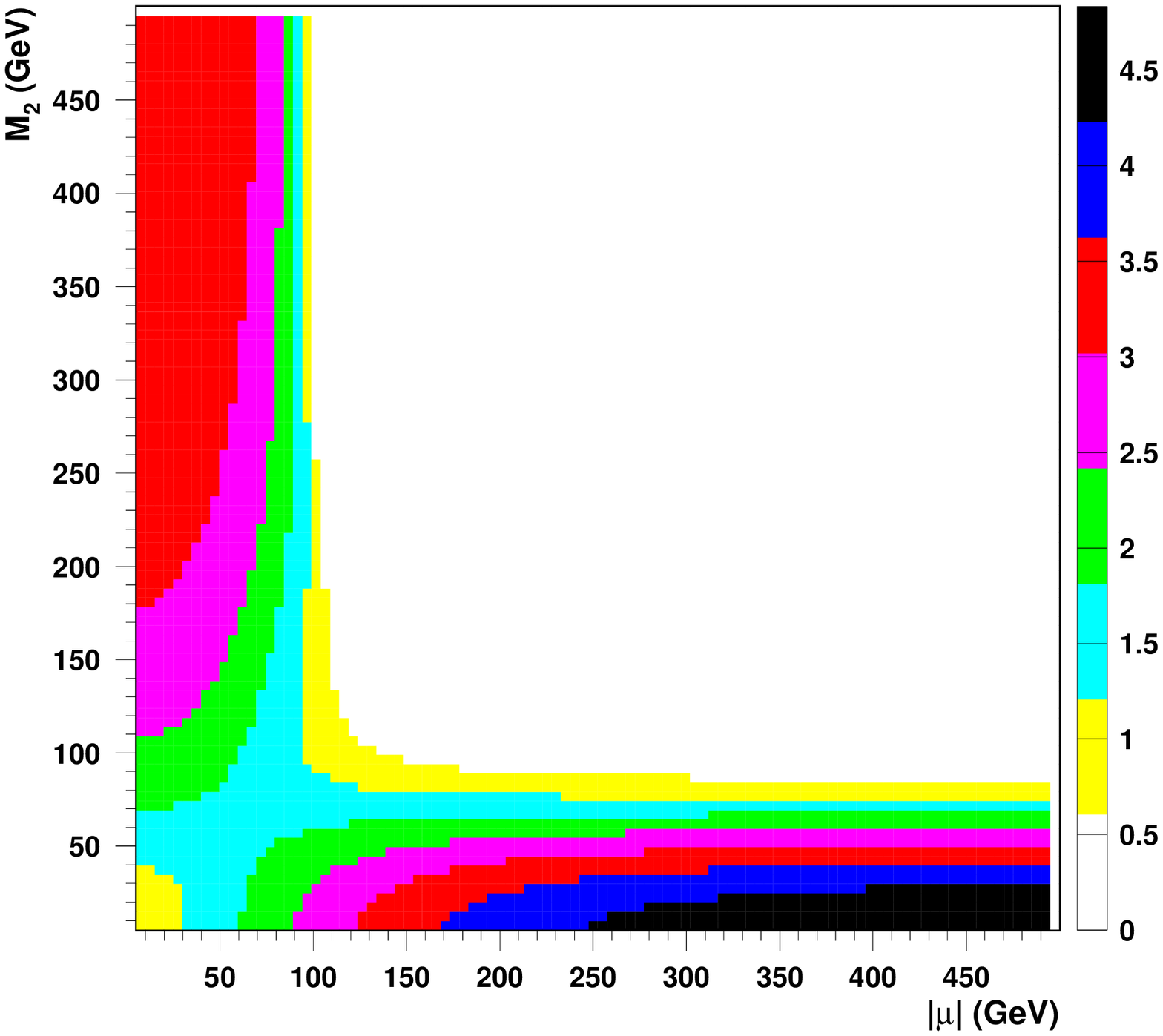}
\includegraphics[width=0.48\linewidth]{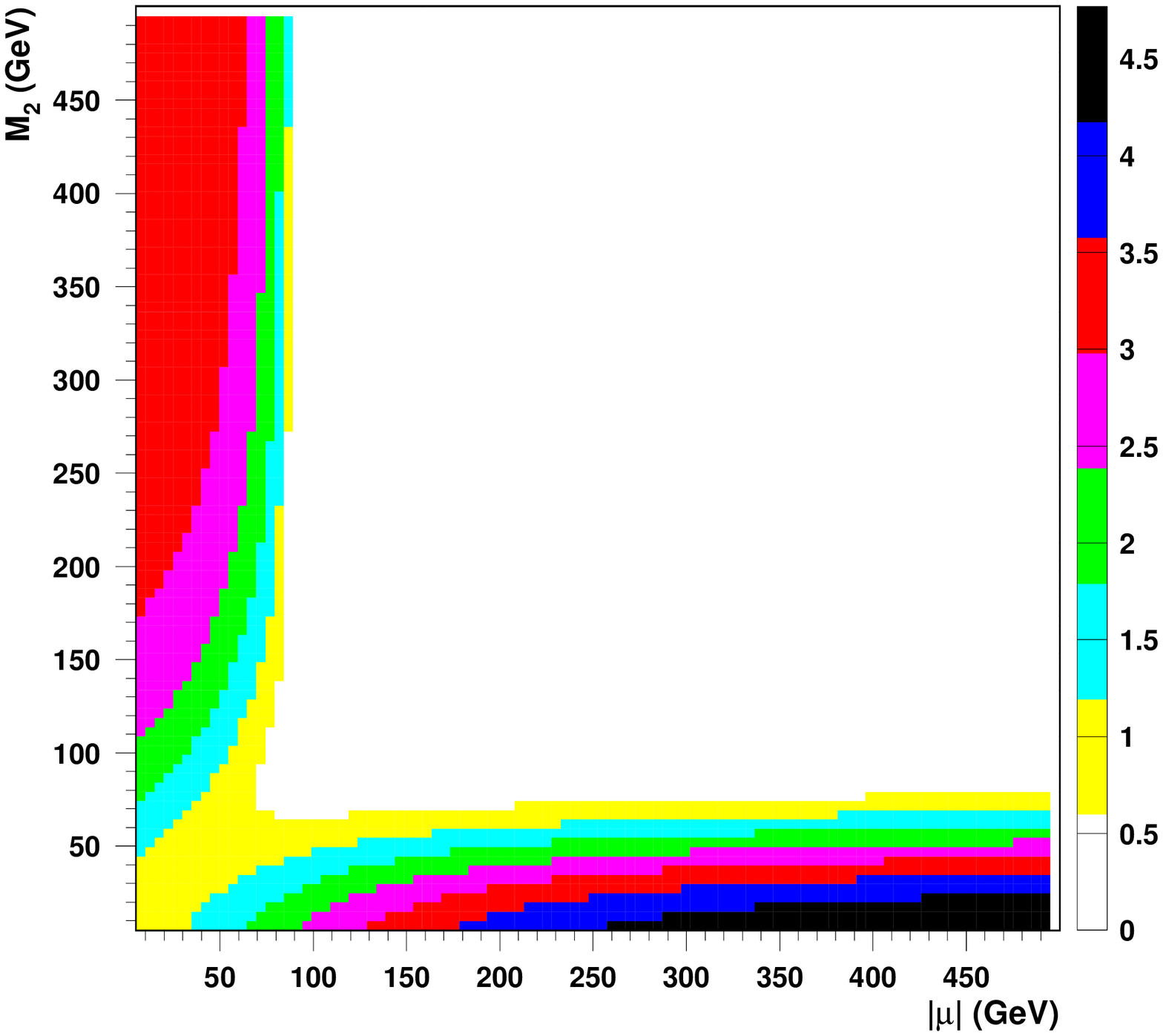}
\includegraphics[width=0.48\linewidth]{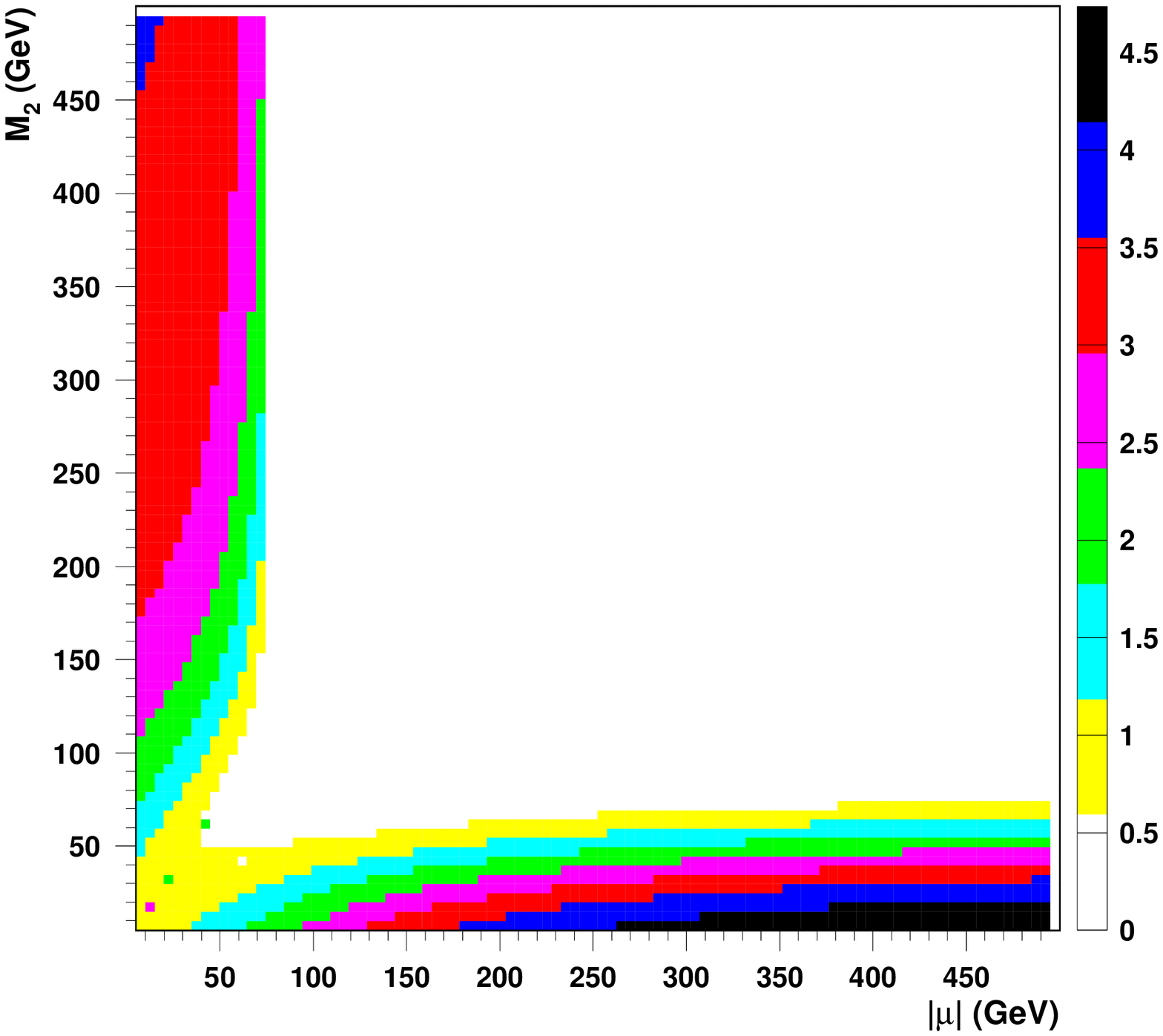}
\caption{Contours of the constant chargino production cross section on
the $(M_2,|\mu|)$ plane for $\tan\beta=1.5$, sneutrino mass $70$ GeV
and $\phi_{\mu}=0,\pi/2,3\pi/4,\pi$.}
\label{fig:crossdist}
\end{center}
\end{figure}

Figure~\ref{fig:crossdist} shows the cross-section distributions in
the ($M_2,|\mu|$) plane for chosen values of the phase and a low value
of $\tan\beta$ and sneutrino mass of 70 GeV.

\begin{figure}[htbp]
\begin{center}
\includegraphics[width=0.48\linewidth]{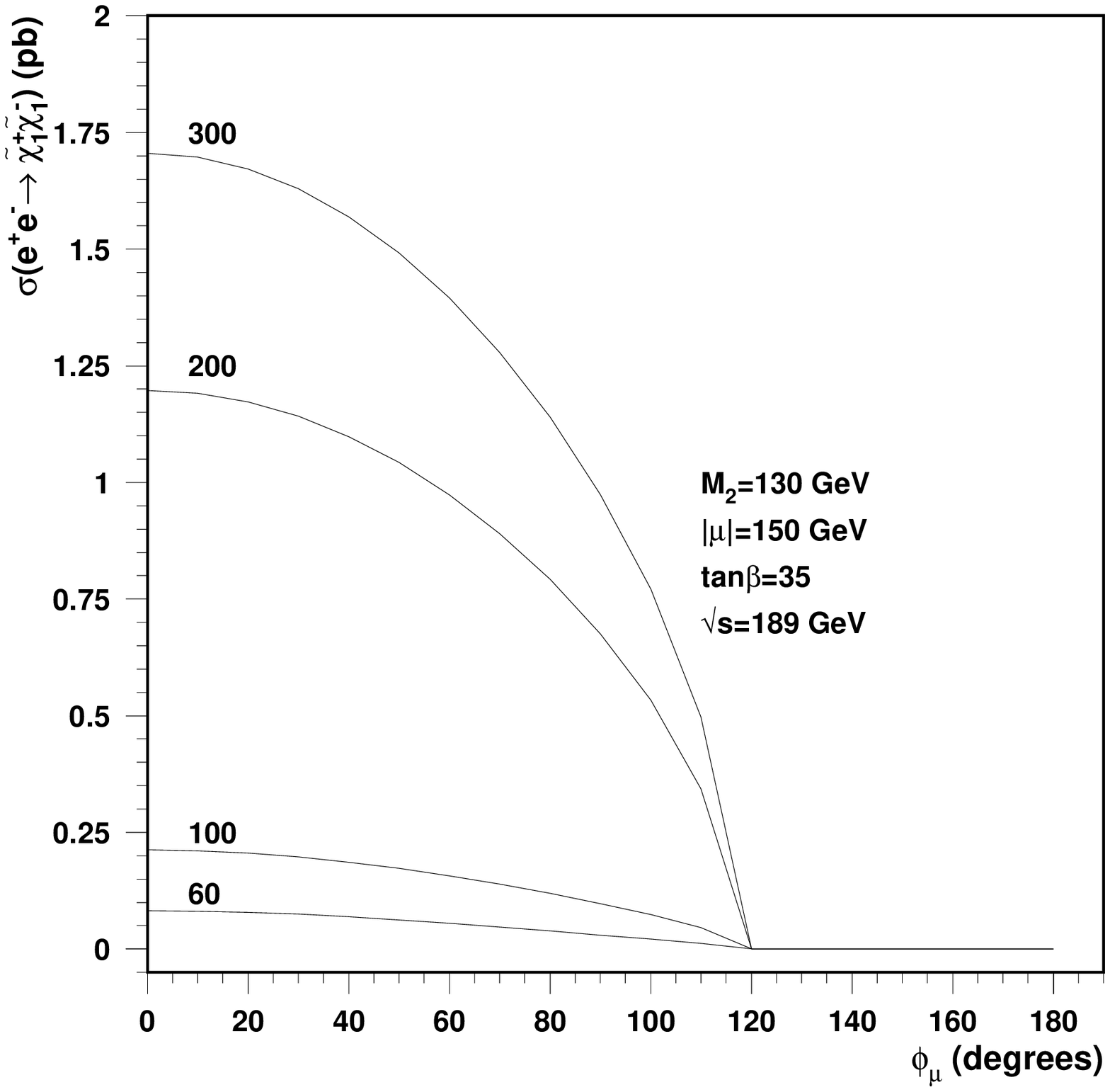}
\includegraphics[width=0.48\linewidth]{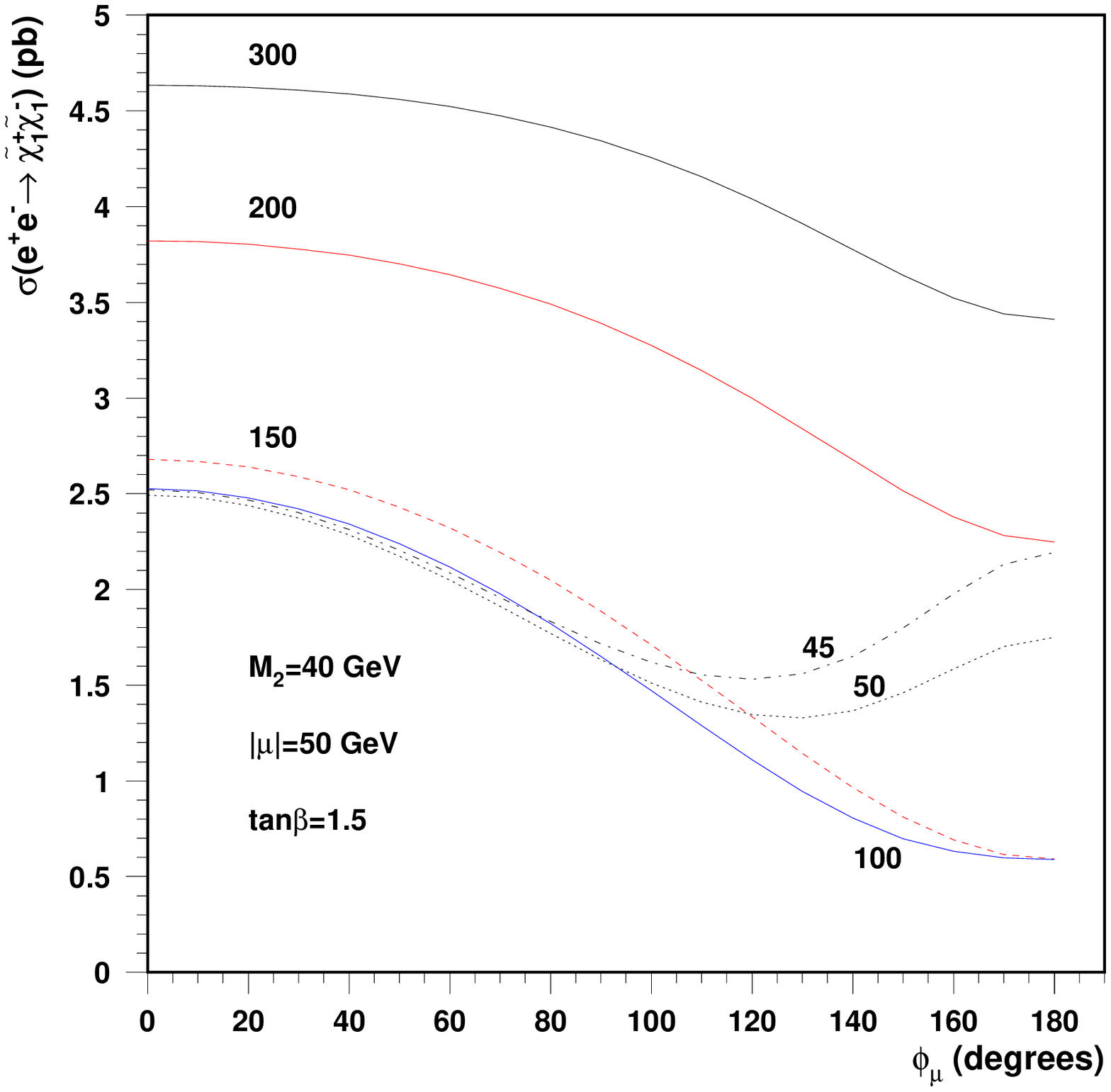}
\caption{Cross section for the chargino production as a function of
$\phi_{\mu}$ for two chosen sets of MSSM parameters and several
sneutrino masses (marked close to the corresponding curves).}
\label{fig:xsec}
\end{center}
\end{figure}

As expected~\cite{CHOI} the chargino production cross section depends
on $\phi_{\mu}$ in most cases through the kinematical effects, i.e.
through the chargino mass dependence on $\phi_{\mu}$.  One can for
instance see in the left plot of fig.~\ref{fig:xsec} a strong phase
space dependence of the cross section,
as the chargino mass increases from below to above the kinematical
threshold.  Nevertheless, the couplings involving the sneutrino also
depend on $\mu$ phase, and therefore one could observe some important
non-kinematical dependencies.

A particularly interesting, from the point of view of this paper, is
the case where the minimal cross section does not occur for one of the
two real $\mu$ values, examined by LEP, but it is reached for some
$\mu$ phase between $0$ and $\pi$.  This is illustrated with the right
plot of figure~\ref{fig:xsec}.  One can see that for higher values of
the sneutrino mass, e.g. greater than 100 GeV, the minimal cross
section is obtained, as could be expected, for real negative $\mu$.
But once one considers small sneutrino masses, this does not hold
anymore and one finds minimal cross section for complex $\mu$.  It can
be a potential loophole for the LEP limits.

\begin{figure}[t]
\begin{center}
\includegraphics[width=0.7\linewidth]{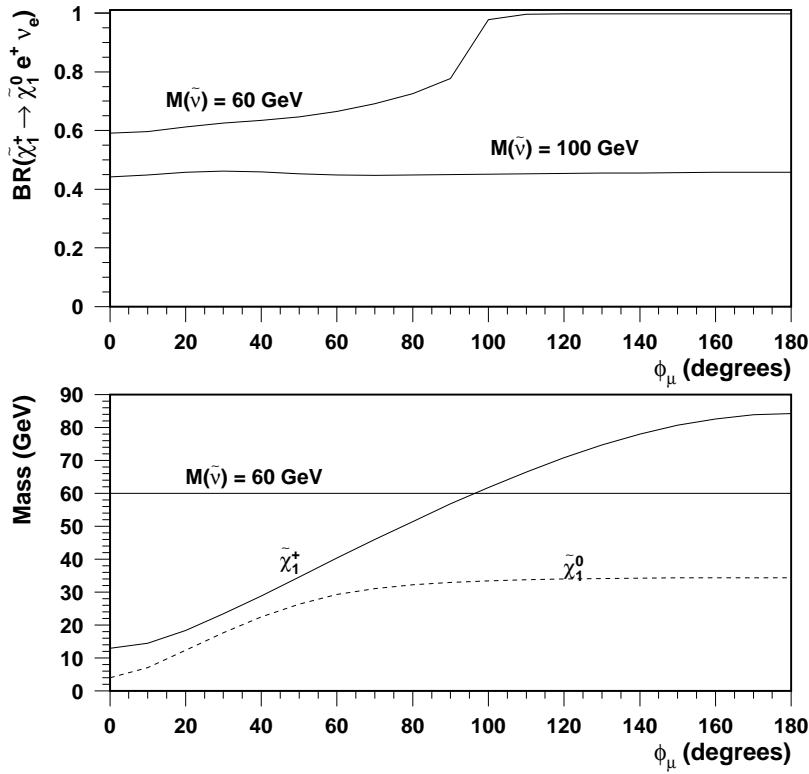}
\caption{Branching ratios for leptonic chargino decays as a function
of $\phi_{\mu}$, for $\tan\beta=1.5$, $M_2=60$ GeV, $|\mu|=140$ GeV
and several values of the sneutrino mass.}
\label{fig:br}
\end{center}
\end{figure}

The branching ratios do not show any dramatic effects. For high
$\tan\beta$ and for low $\tan\beta$ and high sneutrino masses they are
very weakly dependent on the $\mu$ phase.  For low $\tan\beta$ and low
sneutrino masses the branching ratios dependence on $\phi_{\mu}$ can
be explained by the increase of the chargino mass at large
$\phi_{\mu}$ and therefore the opening of the direct decay channel to
sneutrinos.  This fact is a transcription in the complex parameter
language of the well known fact that for negative $\mu$ one has
enhanced leptonic branching fractions. This is illustrated in
figure~\ref{fig:br}, where the leptonic branching ratio is shown as a
function of $\phi_{\mu}$, in parallel with the chargino/neutralino
mass dependence.

It is a general observation that the branching ratios do not have any
local minima between
$\phi_{\mu}=0$ and $\pi$ and therefore we can neglect for the time
being the different experimental sensitivities to different chargino
pair production signatures (fully leptonic, fully hadronic and
``semileptonic'' decays).  It is still possible that a detailed
comparison of expectations and data for each signature finds weak
points, where more luminosity is needed to actually exclude them for a
given complex phase. But we hope the arguments above are sufficient to
indicate that there is no fundamental problem that an increase of
statistics (the LEP experiments currently have collected more than 10
times the statistics used in this analysis) cannot cure.  We therefore
choose to ignore the detailed development of the individual signatures
for this study.

\section{LEP limits revisited}
\label{sec:leprev}

We saw in the previous subsections three ways by which the
introduction of $\mu$ phase could affect the LEP limits, obtained
under the assumption of real MSSM parameters.  The presence of
non-trivial $\mu$ phase can introduce extra degeneracies between the
chargino and neutralino physical masses or/and new cross section
minima for complex $\mu$. We thus will examine, for each MSSM point,
whether the introduction of a new phase either increases the
chargino/neutralino degeneracy or suppresses the cross section.

In this section we revisit the exclusions given by the DELPHI
collaboration with the 189 GeV data~\cite{DELPHI} for a total
integrated luminosity of 158 pb$^{-1}$ collected at this energy.  In
the experimental analysis six mass windows have been considered (see
Table~\ref{tab:masswin}) for 76 MSSM analysis points\footnote{We wish
  to thank here T.~Alderweireld for providing us with the experimental
  data for each point used in the DELPHI paper.}.  This experimental
analysis is not sensitive to mass differences between the chargino and
the neutralino below 3 GeV.  We therefore consider the points with a
mass degeneracy below 3 GeV, for either real or complex $\mu$, as not
excluded.  There exist LEP analyses~\cite{DEGCHARG}, addressing this
problem, and studying the chargino production down to very low mass
differences, but their inclusion is beyond the scope of this paper.

\begin{table}[htbp]
\begin{center}
\begin{tabular}{|c|l|}
\hline
\multicolumn{2}{|c|}{$M_{\chi^+_1} - M_{\chi^0_1}$ regions}\\
\hline
 1 & $\ 3\leq M_{\chi^+_1} - M_{\chi^0_1} <  \ 5\ \mbox{GeV}$ \\
\hline
 2 & $\ 5\leq M_{\chi^+_1} - M_{\chi^0_1} < 10\ \mbox{GeV}$ \\
\hline
 3 & $10\leq M_{\chi^+_1} - M_{\chi^0_1} < 25\ \mbox{GeV}$ \\
\hline
 4 & $25\leq M_{\chi^+_1} - M_{\chi^0_1} < 35\ \mbox{GeV}$ \\
\hline
 5 & $35\leq M_{\chi^+_1} - M_{\chi^0_1} < 50\ \mbox{GeV}$ \\
\hline
 6 & $50\leq M_{\chi^+_1} - M_{\chi^0_1} $ \\
\hline
\end{tabular}
\caption{Mass windows used in the interpretation of chargino searches
based on the $\sqrt{s}=189$ GeV DELPHI data~\cite{DELPHI}.}
\label{tab:masswin}
\end{center}
\end{table}

We first scanned the MSSM parameters, listed in the beginning of this
section, assuming $\mu$ to be real. We compared the theoretical
predictions for the cross-sections with the experimental sensitivity.
\begin{figure}[htbp]
\begin{center}
\begin{tabular}{p{0.474\linewidth}p{0.474\linewidth}}
\mbox{\epsfig{file=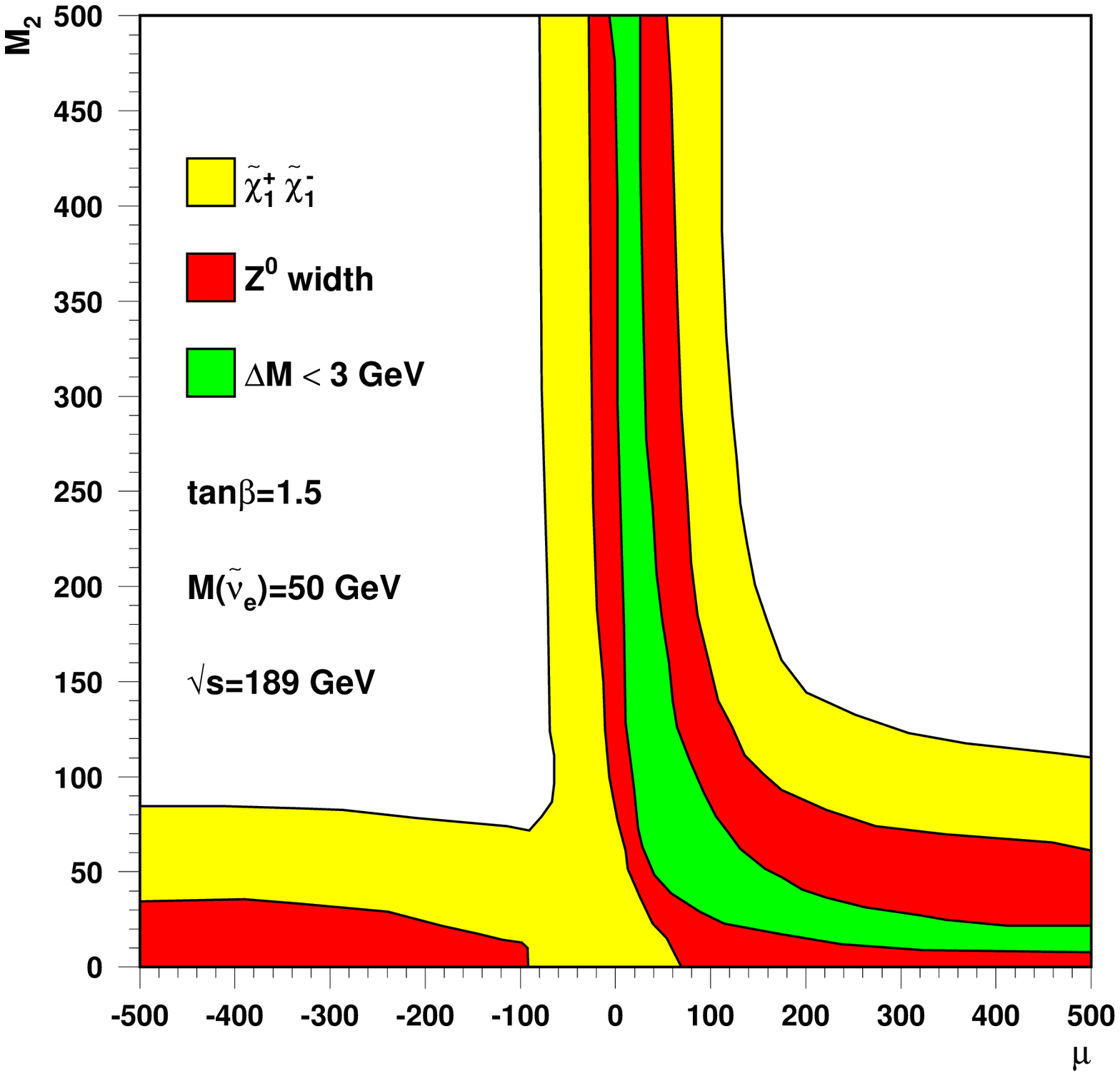,width=\linewidth}}&
\mbox{\epsfig{file=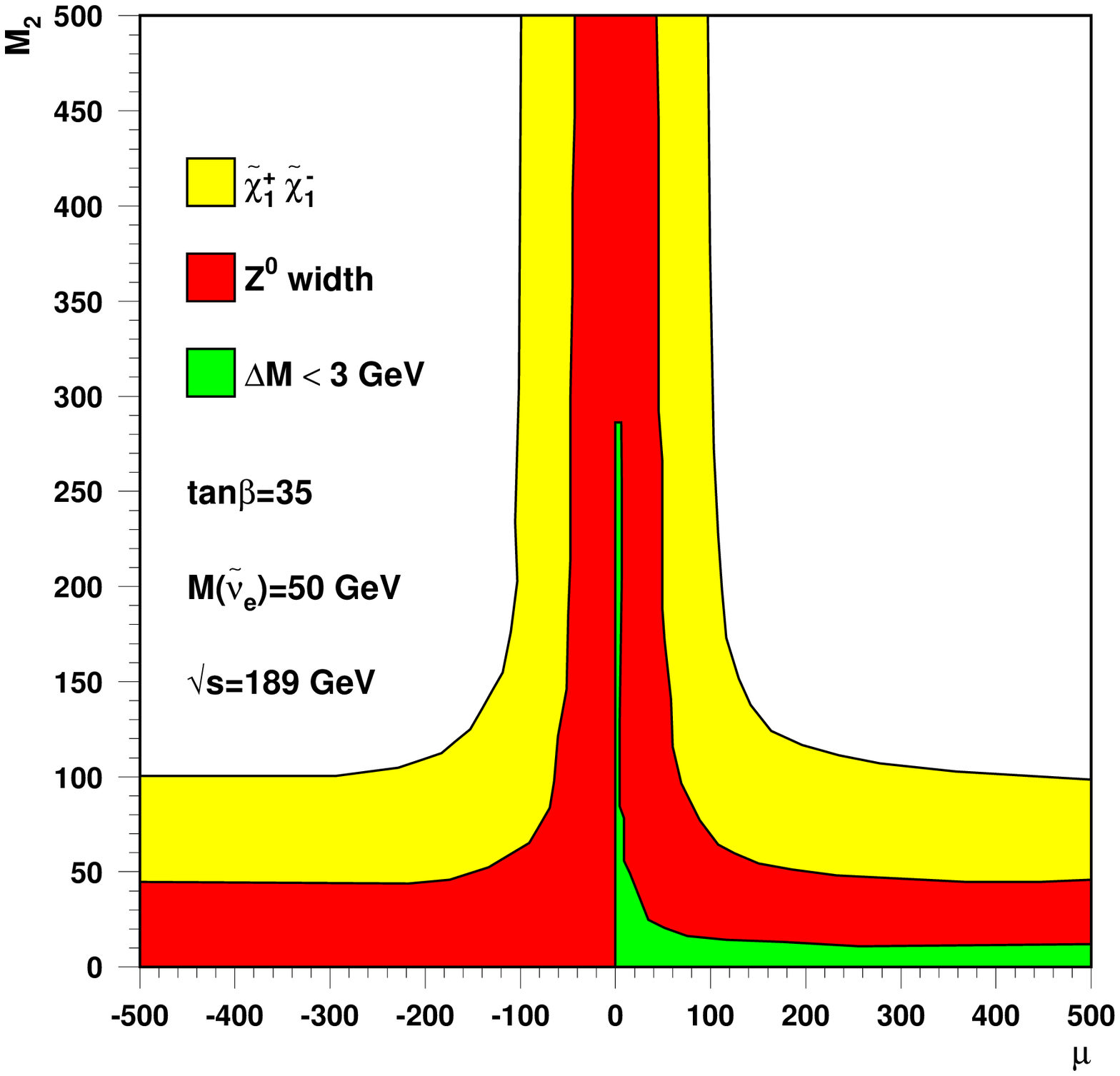,width=\linewidth}}
\end{tabular}
\caption{Excluded areas on the $(M_2$,$\mu)$ plane for the sneutrino
mass 50 GeV (assuming $\mu$ to be real).
\label{fig:realmuexcl}}  
\end{center}
\end{figure}
Figure~\ref{fig:realmuexcl} shows the excluded range in the
$(M_2,\mu)$ plane, consistent with the one published in \cite{DELPHI}.
The inclusion of $\mu$ phase, makes this ``standard'' way of
presenting the excluded area obsolete. One does not have anymore two
possible values for the phase of $\mu$ ($0$ and $\pi$), but a
continuous spectrum. We need therefore to present the exclusions in
the $(M_2,|\mu|)$ plane. The excluded range of
figure~\ref{fig:realmuexcl}, obtained under the assumption of the
reality of $\mu$, has to be ``folded'' around the $M_2$ axis, so that
the new region consists of only the points excluded, at a given $M_2$,
for {\em both} $\phi_\mu=0$ and $\phi_\mu=\pi$, i.e. is given by the
common part of regions excluded for the two extreme $\mu$ phase
values. Actually, it essentially coincides with the excluded range
obtained for real negative $\mu$, ``reflected'' around the $M_2$ axis.
We also report on the same figure, with a different (dark grey) color,
the regions where the point is excluded for only one, negative or
positive, value of $\mu$.

We then check, scanning over the phase of $\mu$, with a step $\pi/18$,
whether the excluded area obtained in this way remains valid for any
$\phi_{\mu}$.  Simultaneously we scan also over the different values
of sneutrino mass and over $\tan\beta$.

As is also custom in "real-parameter" analyses, we add the constraints
coming from the $Z^0$ decay width into unknown particles.  This is of
great help, mostly for regions of high degeneracy.  The experimental
limit used, is
\cite{moenig}: %
\begin{eqnarray}
\Gamma_{new} \leq  6.4 \mathrm{MeV~at~95\%~C.L.}
\label{eq:zwidth}
\end{eqnarray}
The value~(\ref{eq:zwidth}) can be translated into an upper limit on
the total cross section associated with the production of new
particles (in our case charginos and neutralinos) at $\sqrt{s}=M_Z$.
The maximal allowed cross section for supersymmetric particle
production is $\sigma_{new} \leq 152$ pb at $95$\% C.L.  The
contributions are of course calculated with the formulae derived for
the complex MSSM parameters.

\begin{figure}[htbp]
\begin{center}
\begin{tabular}{p{0.474\linewidth}p{0.474\linewidth}}
\mbox{\epsfig{file=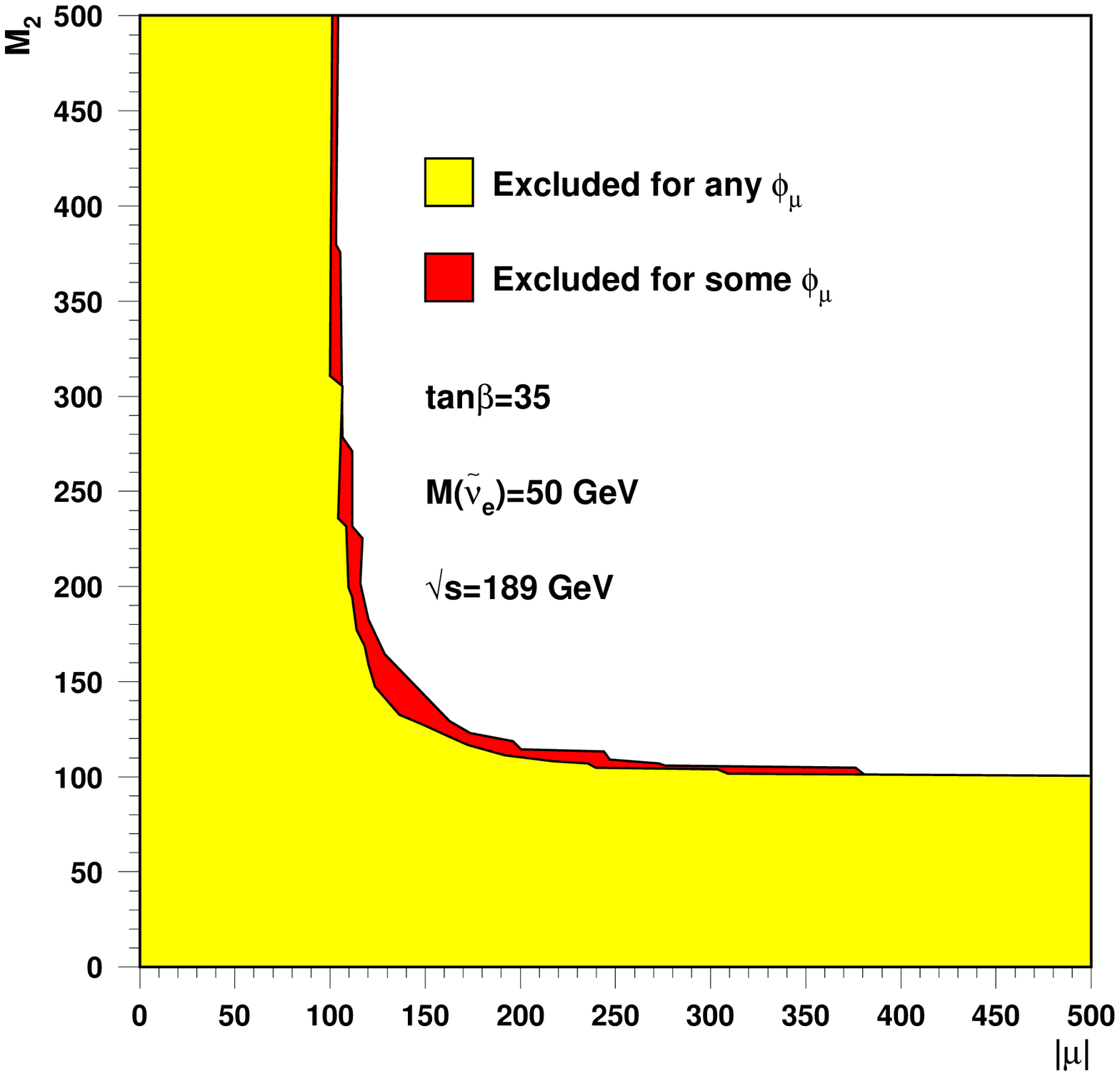,width=\linewidth}}
&
\mbox{\epsfig{file=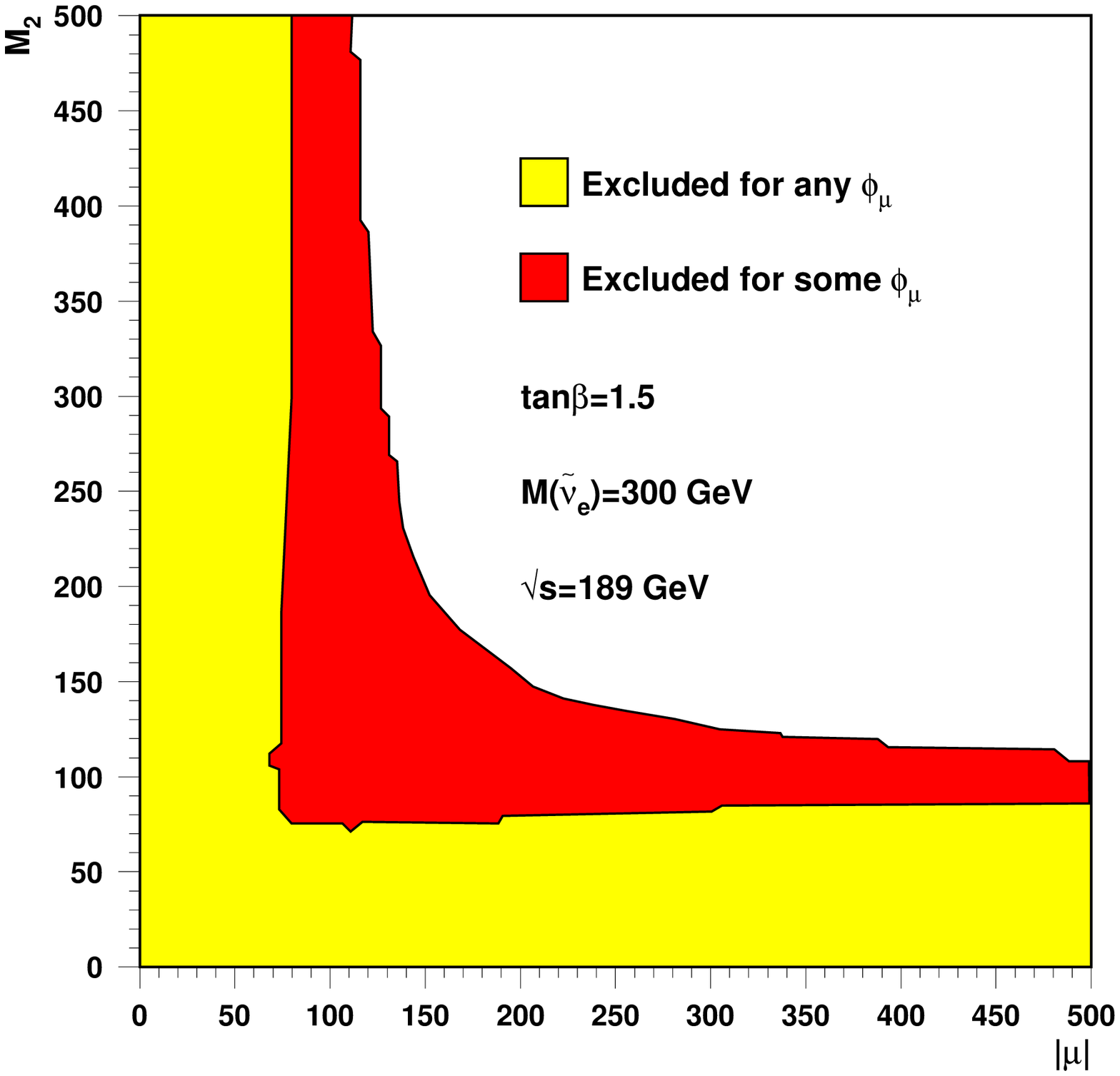,width=\linewidth}}
\\
\caption{Excluded areas in the ($M_2,|\mu|$) plane for large
$\tan\beta$.}
\label{fig:excltb35}
&
\caption{Excluded areas in the ($M_2,|\mu|$) plane for small
$\tan\beta$ and heavy sneutrino.}
\label{fig:excltb1.5_300}
\end{tabular}
\end{center}
\end{figure}
For large $\tan\beta$, the excluded region remains robust, and fully
connected, for any sneutrino masses.  This result is illustrated with
figure~\ref{fig:excltb35}.  The same is true for small values of
$\tan\beta$ and high sneutrino masses, as can be seen in
figure~\ref{fig:excltb1.5_300}.

\begin{figure}[htbp]
\begin{center}
\includegraphics[width=0.48\linewidth]{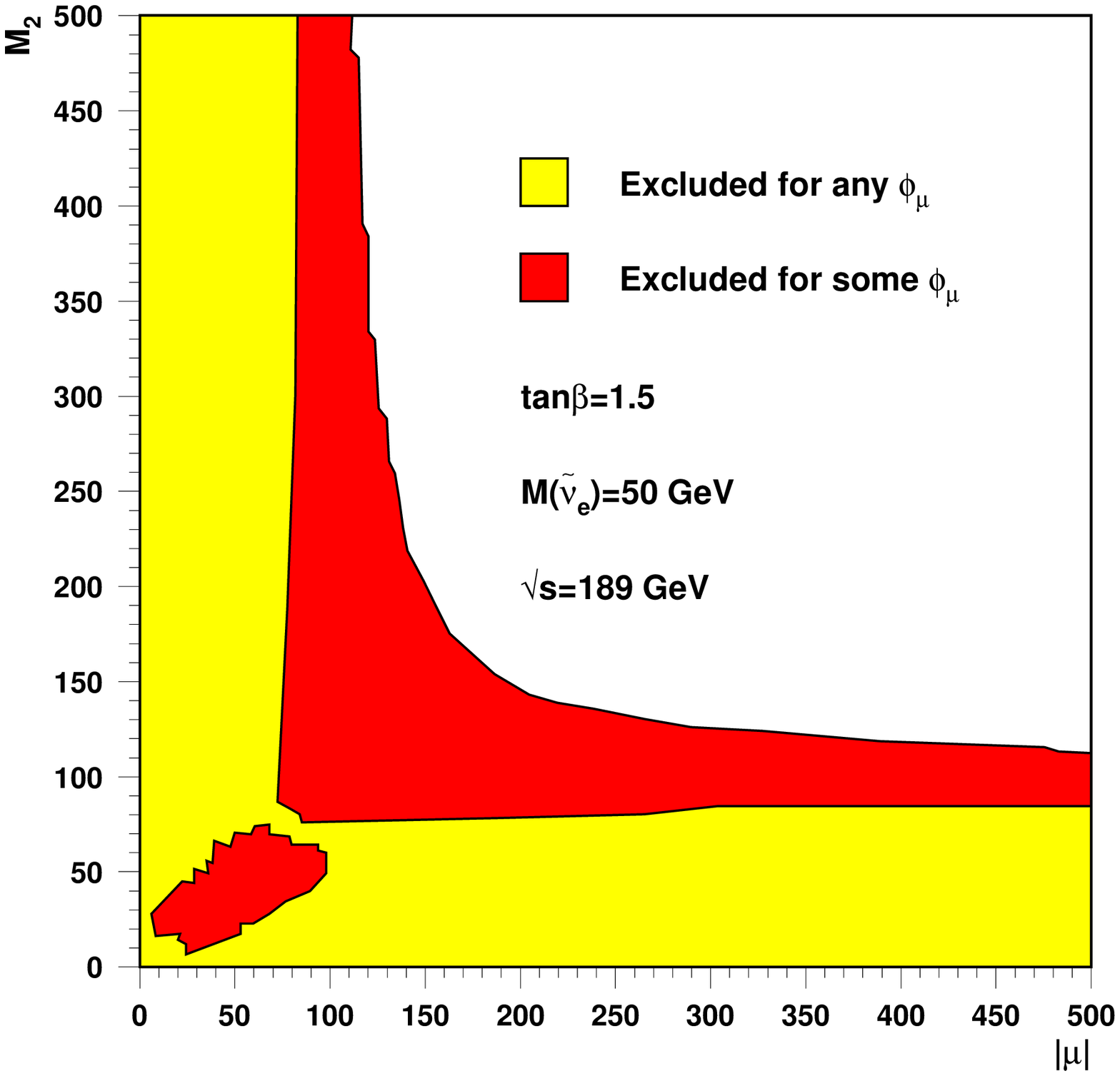}
\includegraphics[width=0.48\linewidth]{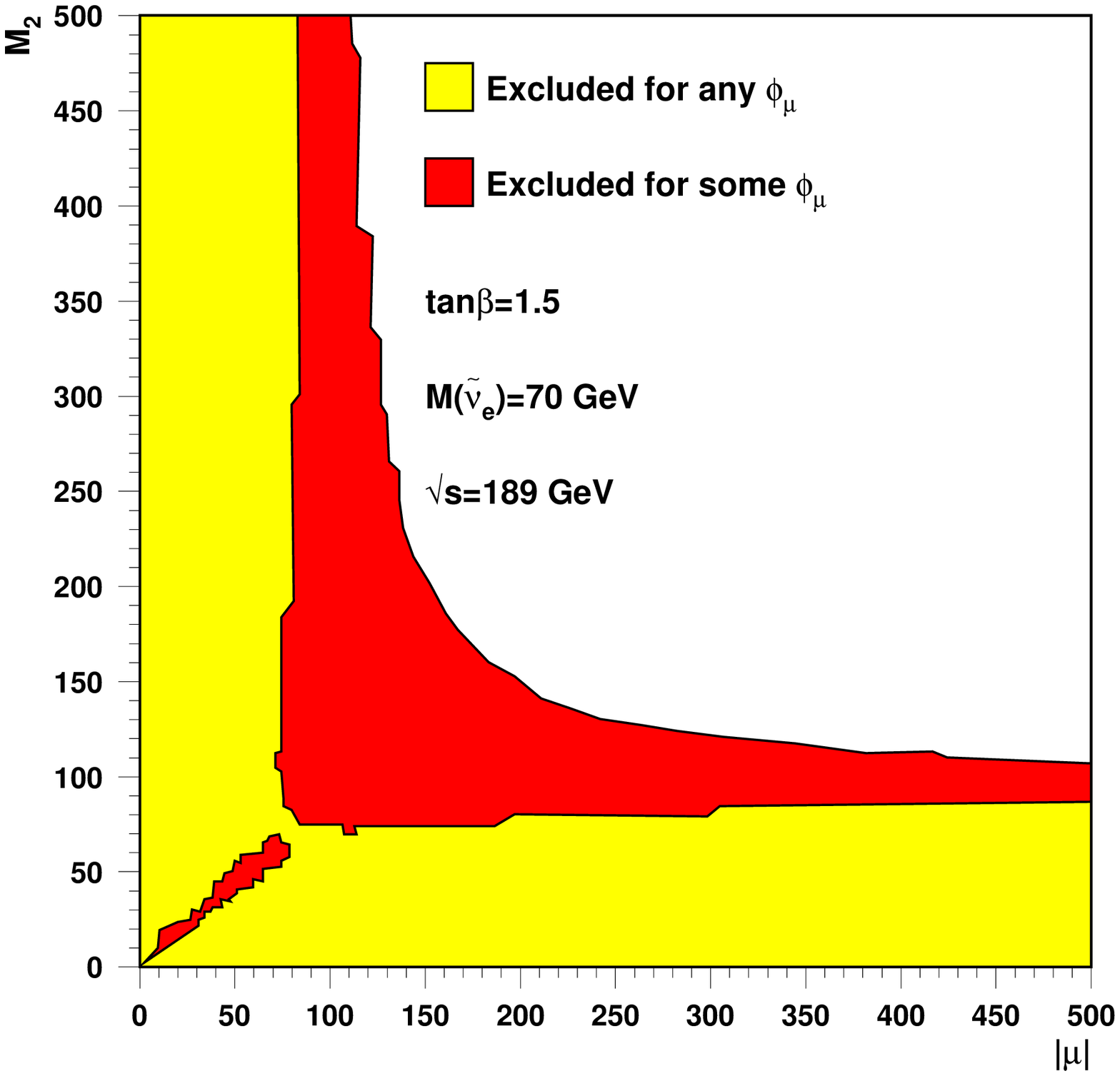}
\caption{Excluded areas in the ($M_2,|\mu|$) plane for small
$\tan\beta$ and light sneutrino.}
\label{fig:excltb1.5_70}
\end{center}
\end{figure}

\begin{figure}[htbp]
\begin{center}
\includegraphics[width=0.48\linewidth]{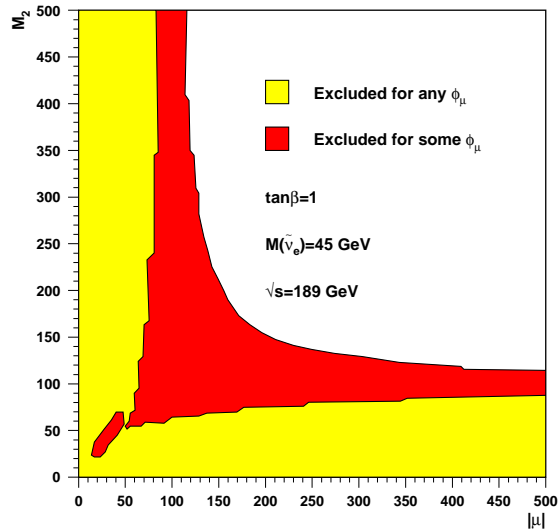}
\caption{Excluded areas in the ($M_2,|\mu|$) plane for
$\tan\beta = 1$ and light sneutrino.}
\label{fig:excltb1_45}
\end{center}
\end{figure}

The situation is more complicated for small values of $\tan\beta$ and
low sneutrino masses. For low sneutrino masses, in the range 50-80
GeV, one finds large areas not excluded anymore inside the region that
is normally excluded for real value $\mu$'s. This is illustrated by
figure~\ref{fig:excltb1.5_70}.  One can remark that for these areas
$M_2$ and $|\mu|$ are connected by the approximate relation $M_2 \sim
\mu \tan\beta$. It is the well known low sensitivity region,
problematic even for real parameter analyses, and concerns the $\mu$
phases close to $\pi$.  The landscape becomes a little more
complicated for $\tan\beta$ exactly equal to 1, where as was shown in
the previous section, there is a new high degeneracy region developing
around $|\mu|=70$ GeV. Figure~\ref{fig:excltb1_45}, shows these two
types (low cross section and high degeneracy) of not-excluded regions
for $\tan\beta=1$ and sneutrino mass of 45 GeV.  Here, we have to
stress once more, that the Higgs searches, analyzed under the
assumption of complex phases \cite{KANE2}, have most probably excluded
the value of $\tan\beta=1$. The analysis of \cite{KANE2} having an
indicative character, when repeated by the LEP working groups, could
firmly establish the complementary role of Higgs to chargino searches
in what regards phases.


Hence, as a first conclusion, we can say that, taking into account the
non-trivial phase of the $\mu$ parameter in the chargino searches at
LEP may lead to the appearance (or the enlarging, compared to the
standard analysis) of unexcluded areas in the $(M_2,|\mu|)$ plane.
These are:
\begin{itemize}
\item[a)] A pronounced unexcluded region for light sneutrino
  $m_{\tilde{\nu}}\sim 50-70$ GeV and small value of $\tan\beta$. This
  is due to local minima of the cross sections, and can be cured by a
  simple increase in statistics. Indeed the real-parameter exclusions
  can be restored if we arbitrarily scale by a factor 9 the statistics
  of the data-sample used. The present data-sample by all 4 LEP
  experiments, corresponds certainly to more than 10 times the
  luminosity than the one used here.
\item[b)] An unexcluded region around $\phi=\pi/2$, for purely
  imaginary $\mu$, and $\tan\beta=1$ which is due to an increase of
  the degeneracy region, and the subsequent lowering of the
  experimental efficiencies. Below certain chargino-neutralino mass
  differences (3 GeV), the analysis used here becomes completely
  inoperative. Special low-degeneracy analyses \cite{DEGCHARG} have to
  be used in order to restore these exclusion areas.  Further, a Higgs
  analysis including MSSM parameter phases, of the type performed in
  ref.~\cite{KANE2}, helps to exclude these remaining points, in good
  complementarity to direct chargino searches.
\end{itemize}

\section{The EDM constraints}
\label{sec:zedm}

Further constraints may be obtained from the measurements of the
electric dipole moments of the electron and
neutron~\cite{EDM_E_EXP,EDM_N_EXP}.  They are, however, not model
independent and require more detailed discussion.

We take into account bounds on the MSSM parameter phases given only by
the electron EDM measurement.  This is for two reasons.  First, the
electron EDM depends exactly on the same set of parameters as the
cross sections for the chargino and neutralino production, so we do
not need to introduce any additional variables in our scan.  Second,
the theoretical calculation of the neutron EDM is prone to significant
QCD uncertainties (see e.g. discussion in~\cite{PRS}), so the limits
obtained from its measurements are less established that those given
by the electron EDM.

The limits on $\phi_{\mu}$ coming from the electron EDM measurements
also cannot be treated as absolute.  In a good approximation, the
formulae for the electron EDM can be written down as:
\bea
d_e &\approx& d_1\mathrm{Im}(\mu M_1)\tan\beta + d_2\mathrm{Im}(\mu
M_2)\tan\beta + d_3\mathrm{Im}(A_e M_1^{\star})
\label{eq:edm_exp}
\eea
where coefficients $d_1, d_2, d_3$ depend only on the absolute values
of $|M_1|, |M_2|, |\mu|$ and the slepton and sneutrino masses.  If
$M_1$ and $M_2$ are GUT-related and thus can be both chosen to be
real, eq.~(\ref{eq:edm_exp}) reduces to:
\bea
d_e \approx d_{\mu}\mathrm{Im}(\mu)\tan\beta + d_A\mathrm{Im}(A_e)
\label{eq:edm_expa}
\eea
where for the typical choices of the mass parameters (like the ones we
used in our scan) $|d_{\mu}|/|d_A|\sim {\cal O}(10)$, so that $d_e$ is
significantly more sensitive to $\phi_{\mu}$ than to $\phi_{A_e}$,
particularly for large $\tan\beta$.

The left-right selectron mixing parameter $A_e$ enters formally the
expression for the neutralino production cross section, but it is
multiplied there by the electron mass and can be neglected unless
$A_e$ is really huge, $|A_e|/m_{\tilde{e}}> {\cal O}(10^5)$.  Such
large values, although not excluded by any experimental measurement,
are highly unlikely for theoretical purposes.  Therefore, production
cross sections of the charginos and neutralinos depend effectively
only on the $\mu$ parameter phase, whereas the electron EDM depends on
both $\phi_{\mu}$ and $\phi_{A_e}$.  This leaves a possibility of
cancellation between phases: for any value of $\mathrm{Im}(\mu)$ one
can find a matching value of $\mathrm{Im}A_e$ such that both terms in
eq.~(\ref{eq:edm_expa}) are almost equal and opposite in sign, so that
the electron EDM value predicted by MSSM is below the current
experimental bounds.  However, as mentioned already in the
introduction, such cancellations seems to be entirely accidental (from
the point of view of the electroweak scale physics at least) and
require strong fine tuning between phases and mass parameters - for
light supersymmetric spectrum their values must be correlated with
accuracy ${\cal O}(10^{-2})$~\cite{PRS}.  Nevertheless, their presence
implies that in order to put bounds on the allowed values of
$\phi_{\mu}$ one needs to apply further assumptions.

\begin{figure}[htbp]
\begin{center}
\includegraphics[width=0.5\linewidth]{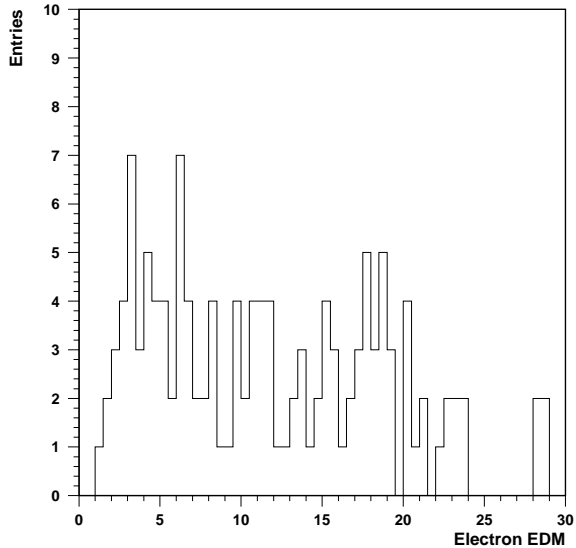}
\caption{Values of the electron EDM calculated for each of the
unexcluded points in ($M_2,|\mu|$) plane, assuming
$A_e=0$, normalized to (divided by) the experimental bound.}
\label{fig:edm_mu}
\end{center}
\end{figure}

First, one may naturally assume that the strong fine-tuning between
the MSSM parameters required for the cancellations in the electron EDM
does not occur.  In this case, ``generic'' limits on the $\mu$ phase
may be obtained assuming that one can neglect the $\mathrm{Im}(A_e)$
term in the $d_e$ expression, setting it to $\mathrm{Im}(A_e)=0$.  We
computed the electron EDM values for the points not excluded by the
chargino searches analysis and plotted them in
figure~\ref{fig:edm_mu}.
As can be seen from fig.~\ref{fig:edm_mu}, all points are excluded by
the condition $|d_e|/d_e^{exp}\leq 1$.  Thus, assuming no
cancellations between the phases in the electron EDM, the excluded
area of the $(M_2,|\mu|)$ plane is not smaller than the one obtained
for real negative $\mu$.

\begin{figure}[htbp]
\begin{center}
\includegraphics[width=0.5\linewidth]{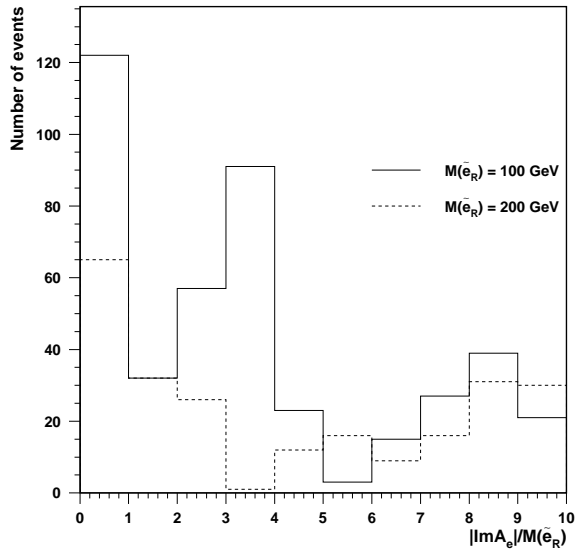}
\caption{Minimal values of $|\mathrm{Im}A_e|/m_{\tilde{e}}$ necessary
  for phase cancellation in the electron EDM, calculated for each of
  the unexcluded points in ($M_2,|\mu|$) plane, for two choices of the
  right selectron mass.}
\label{fig:amin}
\end{center}
\end{figure}

Second, one may decide to take into account the possibility of
cancellations between the phases in the electron EDM.  Then, for most
of the mass parameter choices, the relative amplitude of the
coefficients $d_{\mu}$ and $d_A$ in eq.~(\ref{eq:edm_expa}) implies
that substantial $\mu$ phase requires also large value of the
imaginary part of the $A_e$ parameter, $\mathrm{Im}(A_e)\sim {\cal
O}(10)\times \tan\beta \times \mathrm{Im}(\mu) $, to keep the full
electron EDM below the experimental bound.  In figure~\ref{fig:amin}
we plot, calculated for each of the unexcluded points on $(M_2,|\mu|)$
plane, the minimal $|\mathrm{Im}(A_e)|$ values required to make
$\phi_{\mu}-\phi_{A_e}$ cancellation in the electron EDM possible.  As
can be seen from the figure, they depend on the right selectron mass,
the larger it is the higher $A_e$ are required\footnote{The ``peak
structure'' visible in the plot is artificial and depends on density
of scan over MSSM parameters.}.

The maximal value of $A_e$ parameter could be constrained if one
assumes unification of all LR mixing parameters (both in slepton and
squark sectors) at the GUT scale.  In such a case, LR mixing parameter
in the stop sector should not be larger than $|A_t|/m_{\tilde{t}}\sim
\sqrt{3}$, in order to avoid color symmetry breaking~\cite{MUNOZ} and,
on the base of the unification assumption, a similar limit may be
applied to $A_e$.  As can be seen from figure~\ref{fig:amin}, even if
we allow for $\phi_{\mu}-\phi_A$ cancellations, a cut on
$|\mathrm{Im}A_e|/m_{\tilde{e}}<2-3$ eliminates a large fraction of
the unexcluded points on ($M_2,|\mu|$) plane for right selectron mass
of the order of 100 GeV, just above the current experimental bound,
and this fraction grows quickly with $m_{{\tilde e}_R}$.

Finally, we comment on the possibility of a non-vanishing phase of the
$M_1$ parameter.  Such a phase, even if not explicitly present in the
chargino mass matrix and expressions for the production cross section,
affects the interpretation of chargino searches, through its influence
on the decay rates to neutralinos, and through the increased freedom
for cancellations in the electron EDM, as obvious from the form of
eq.~(\ref{eq:edm_exp}) (however, as shown in~\cite{PRS}, for light
SUSY spectrum such cancellation again require precise fine-tuning
between parameter values).  Varying $M_1$ phase one changes the
physical masses of neutralinos and thus the size of mass splitting
between the lightest neutralino and chargino, affecting the
experimental efficiency of searches.  Hence, the possible impact of
the additional phase could be very important.  However, from the
theoretical point of view, allowing for different phases of $M_1$ and
$M_2$ makes sense most likely only if one rejects the assumption of
gaugino mass unification, i.e.  simultaneously gives up the relation
connecting absolute values of masses, $|M_1| = \frac{5}{3}
\tan\theta_W^2|M_2|$.  In this case $|M_1|$ becomes an additional free
parameter and scanning over it one can always find values for which
lightest neutralino-chargino mass splitting tends to vanish, so that
``standard'' chargino searches cannot discover it.  Special
experimental strategies concerning this experimental ``blind spot''
have been developed, as mentioned above~\cite{DEGCHARG}.  These
searches, being based on the detection of the SUSY particles through
the extra radiation of an initial state photon~\cite{GUNION}, need the
full LEP2 luminosity to become as sensitive as the standard ones, and
they will eventually reach close to the kinematical limits only after
the end of LEP.  Therefore, any meaningful analysis not assuming GUT
unification of gaugino masses, i.e. free complex $M_1$ parameter, must
take into account these searches, has to await for the end of LEP and
is beyond the scope of this paper.

\section{Conclusions}
\label{sec:conclusions}

In conclusion, we recalculated the production and decay rates of
charginos and neutralinos at LEP, in the case of complex MSSM
parameters.  We performed an extensive scan over the relevant MSSM
parameters and compared the expected signals to data from DELPHI,
taken at 189 GeV.  We extend the standard LEP analyses by scanning
also over the $\mu$ phase. This is the only new phase to which these
processes are sensitive, if one assumes $M_1,M_2$ unification at the
GUT scale. The inclusion of $\mu$ phase introduces new points of
degeneracy between the chargino and neutralino masses, lowering the
chargino detection efficiency. It can also lead to cross-sections
lower than the ones obtained for real $\mu$ values and mildly affects
the decay branching ratios. We found that the limits obtained by the
experimental collaboration, which do not take into account the
non-trivial $\mu$ phase are in general robust, apart from the case of
low $\tan\beta$ and low sneutrino mass for which unexcluded points
appear in a region around $M_2 \sim \tan\beta |\mu|$.  These points
can be excluded, restoring the real-case limits, if one takes into
account the limits on new physics obtained by the measurement of the
$Z^0$ width and the experimental constraints on electron EDM.  The
latter statement assumes that $\mathrm{Im}(A_e)$ is not precisely
fine-tuned so that it cancels the $\mathrm{Im}(\mu)$ contribution to
the EDM.  Apart, possibly, from cases of extreme degeneracy between
charginos and neutralinos induced by phases, present anyway also in
the real-parameter case, we found no fundamental loophole in the LEP
exclusions that would not be covered by the final LEP luminosity.
Further, the influence of $\mu$ phase is more prominent for low
$\tan\beta$, in a region where Higgs negative searches can give
complementary exclusions, provided they are also made under the
assumption of complex MSSM parameters.

\vskip 5mm 

\noindent {\large \bf Acknowledgments}

\vskip 2mm

\noindent We would like to acknowledge discussions with J.F.~Grivaz, 
G.~Kane, F.~Richard, S.~Pokorski, M.~Winter and many members of the
``CNRS, Groupement De Recherche sur la Supersym{\'e}trie'' as well as
the extremely valuable help of T.~Alderweireld and P.~Rebecchi who
provided us with the DELPHI data.

\vskip 5mm

\renewcommand{\thesection}{Appendix}
\renewcommand{\theequation}{\Alph{section}.\arabic{equation}}

\setcounter{equation}{0}
\setcounter{section}{0}

\section{Conventions and Feynman rules}
\label{app:lagr}

For easy comparison with other references we spell out our
conventions.  They are similar to the ones used in ref.~\cite{FEYRUL}.
We present only the part of the MSSM Lagrangian which we are
interested in, i.e. electroweak interactions of gauge, Higgs and
slepton supermultiplets.  The MSSM matter fields form chiral
left-handed superfields in the following representations of the $SU(2)
\times U(1)$ gauge group (the generation index is suppressed):

\begin{eqnarray*}
\begin{array}{lll}
\mathrm{Scalar~field}&  \mathrm{Weyl~Fermion~field} & 
SU(2) \times U(1) ~\mathrm{representation}\\
L=\left(\begin{array}{c}\tilde\nu_0\\E\end{array}\right)& 
\mbox{\hskip 1cm} l=\left(\begin{array}{c}{\nu}\\
e\end{array}\right)&
\mbox{\hskip 2cm}(2,-1)\\
E^{c} &\mbox{\hskip 1cm}e^c&\mbox{\hskip 2cm} (0,2)\\
H^{1}\left(\begin{array}{c} H_1^1\\H_2^1\end{array}\right)&
\mbox{\hskip 1cm}\tilde{h}^1\left(\begin{array}{c} \tilde{h}_1^1\\
\tilde{h}_2^1\end{array}\right)& 
\mbox{\hskip 2cm}(2,-1) \\
H^{2}\left(\begin{array}{c} H_1^2\\H_2^2\end{array}\right)&
\mbox{\hskip 1cm}\tilde{h}^2\left(\begin{array}{c} \tilde{h}_1^2\\
\tilde{h}_2^2\end{array}\right)& 
\mbox{\hskip 2cm}(2,1)
\label{eq:superkm}
\end{array}
\end{eqnarray*}
Two $SU(2)$-doublets can be contracted into an $SU(2)$-singlet,
e.g. $H^1 H^2 = \epsilon_{ij}H^1_i H^2_j = - H^1_1 H^2_2 + H^1_2
H^2_1$ (we choose $\epsilon_{12}=-1$; lower indices (when present)
will label components of $SU(2)$-doublets).  The superpotential and the
soft terms are defined as:
\bea
W = Y_e H^1 L E +  \mu H^1 H^2
\label{eq:superpot}
\eea
\bea
{\cal L}_{soft} &=& - M_{H^1}^2 H^{1\dagger} H^1 - M_{H^2}^2 H^{2\dagger} H^2
- L^{\dagger} M_L^2  L 
- E^{c\dagger} M_E^2 E^c\nonumber\\
&+&\left(\f{1}{2} M_2 \tilde{W}^i \tilde{W}^i
              + \f{1}{2} M_1 \tilde{B}\tilde{B}
              + m^2_{12} H^1 H^2
              + Y_e A_e H^1 L E^c 
              + \mathrm{H.c.}\right)
\label{eq:lsoft}
\eea 

where we extracted Yukawa coupling matrices from the definition of the
$A_e$ coefficient.

In general, the Yukawa couplings and the masses are matrices in the
flavor space.  Simultaneous rotation of the fermion and sfermion
fields can diagonalize the Yukawa couplings (and simultaneously
fermion mass matrices), leading to so-called ``super-KM'' basis, with
flavor diagonal Yukawa couplings and neutral current fermion and
sfermion vertices.  We give all the expressions already in the super-KM
basis (see e.g.~\cite{FCNC97} for a more detailed discussion).

The slepton mass matrices in the super-KM basis have the following
form:
\bea
{\cal M}^2_{\tilde{L}} &=&  
\left( \begin{array}{cc}
M^2_L + m_e^2 + \frac{\cos 2\beta}{2}(M_Z^2 - 2M_W^2)\hat{\mbox{\large 1}} & 
-m_e(\tan\beta \mu\hat{\mbox{\large 1}} + A_e^{\star})\\
-m_e(\tan\beta \mu^{\star}\hat{\mbox{\large 1}} + A_e) & 
M^2_E + m_e^2 -\cos 2\beta (M_Z^2 - M_W^2)\hat{\mbox{\large 1}}\\
\end{array}\right)\nonumber\\ 
\nonumber\\
{\cal M}^2_{\tilde{\nu}}& =&
M^2_L +  \frac{\cos 2\beta}{2} M_Z^2\hat{\mbox{\large 1}}
\label{eq:sfmass}
\eea

\noindent where $\theta_W$ is the Weinberg angle and
$\hat{\mbox{\large 1}}$ stands for the $3 \times 3$ unit matrix.

The matrices ${\cal M}^2_{\tilde{\nu}}$ and ${\cal M}^2_{\tilde{L}}$  can be
diagonalized by additional unitary matrices $Z_{\nu}$ ($3\times 3$)
and $Z_L$ ($6\times 6$), respectively
\bea 
\left({\cal M}^2_{\tilde{\nu}}\right)^{diag} = 
Z_{\nu}^{\dagger} {\cal M}^2_{\tilde{\nu}} Z_{\nu}
&\hskip 3cm&
\left({\cal M}^2_{\tilde{L}}\right)^{diag}  =  
Z_L^{\dagger} {\cal M}^2_{\tilde{L}} Z_L 
\label{eq:zdef}
\eea

The physical (mass eigenstates) sleptons are then defined in terms of
super-KM basis fields~(\ref{eq:superkm}) as:
\bea
\tilde{\nu} = Z_{\nu}^{\dagger} \tilde\nu_0
\hskip 3cm
\tilde{L} = Z_L^{\dagger} \left( \begin{array}{c} E \\ E^{c\star} \end{array} 
\right)
\label{eq:sferdef}
\eea

Throughout this paper we assume that the flavor and CP violation due
to the flavor mixing in the sfermion mass matrices is negligible from
the point of view of chargino and neutralino production, i.e. matrices
$M^2_{L,E}, A_e$ are almost diagonal in the super-KM basis, so that
also $Z_{\tilde{\nu}},Z_L \approx \hat{\mbox{\large 1}}$.  However, one
should remember that even such very small values of the $A_e$
parameter, if contain imaginary part, may affect bounds on the $\mu$
phase given by the EDM measurements.

The physical Dirac chargino and Majorana neutralino eigenstates are
linear combinations of left-handed Winos, Binos and Higgsinos
\bea
\chi^+_i = \left(\begin{array}{c}
-iZ_{+}^{i1\star}\tilde{W}^+ + Z_{+}^{i2\star}\tilde{h}^1_2 \\
iZ_{-}^{i1}\overline{\tilde{W}^-} + Z_{-}^{i2}\overline{\tilde{h}^2_1}
\end{array}\right)
\label{eq:chargmass}
\eea
where $\tilde{W}^{\pm} = (\tilde{W}^1\mp\tilde{W}^2)/\sqrt{2}$.
\bea
\chi^0_i = \left(\begin{array}{c}
-iZ_N^{i1\star}\tilde{B}  -iZ_N^{i2\star}\tilde{W}^3 
+ Z_N^{i3\star}\tilde{h}^1_1 + Z_N^{i4\star}\tilde{h}^2_2 \\
iZ_N^{i1}\overline{\tilde{B}} + iZ_N^{i2}\overline{\tilde{W}^3} 
+ Z_N^{i3} \overline{\tilde{h}^1_1} + Z_N^{i4}\overline{\tilde{h}^2_2}
\end{array}\right)
\eea
The unitary transformations $Z^+$, $Z^-$ and $Z_N$ diagonalize the mass
matrices of these fields
\bea
{\cal M}_C = Z_{-}^T \left( \begin{array}{cc} M_2 &
\frac{gv_2}{\sqrt{2}}  \\
\frac{gv_1}{\sqrt{2}} & \mu \end{array} \right) Z_{+}
\eea
and
\bea
{\cal M}_N = Z_N^T \left( \begin{array}{cccc} 
M_1 & 0 & -\frac{g'v_1}{2} &  \frac{g'v_2}{2}\\
0 & M_2 &  \frac{gv_1}{2}  & -\frac{gv_2}{2} \\
-\frac{g'v_1}{2} & \frac{gv_1}{2} & 0    & -\mu \\ 
\frac{g'v_2}{2}& -\frac{gv_2}{2}  & -\mu & 0
\end{array} \right) Z_N
\eea

Using the notation of this Appendix, one can list (in the mass
eigenstate basis) the Feynman rules necessary to calculate cross
sections~(\ref{eq:zcc_diff}),(\ref{eq:zcc})
and~(\ref{eq:znn_diff}),(\ref{eq:znn}):

1) Interactions of charginos and  sneutrinos:\\[5mm]
\begin{tabular}{lp{20mm}l}
\begin{picture}(115,60)(0,0)
\DashArrowLine(50,10)(10,10){6}
\Vertex(50,10){2}
\ArrowLine(50,10)(90,10)
\ArrowLine(50,50)(50,10)
\Text(0,10)[]{$\tilde{\nu}^J$}
\Text(110,10)[]{$(\chi_j^+)^C$}
\Text(50,60)[]{$e^I$}
\end{picture}
&&
\raisebox{10\unitlength}{\begin{tabular}{l}
$ i\left(S_{LC}^{IJi}P_L 
+ S_{RC}^{IJi} P_R \right)$\\ 
\end{tabular}}\\
\end{tabular}

\noindent where
\bea
S_{LC}^{IJi} &=& -g_2 Z_{1j}^+Z_{\tilde{\nu}}^{IJ\star}\\
S_{RC}^{IJi} &=& {m_e^I\sqrt{2}\over v_1} Z_{2j}^{-*}
Z_{\tilde{\nu}}^{IJ\star}
\eea

2) Interactions of charginos and neutralinos with gauge bosons:\\[5mm]
\begin{tabular}{lp{20mm}l}
\begin{picture}(115,60)(0,0)
\Photon(50,10)(10,10){4}{4}
\Vertex(50,10){2}
\ArrowLine(50,10)(90,10)
\ArrowLine(50,50)(50,10)
\Text(0,10)[]{$\gamma^{\mu}$}
\Text(110,10)[]{$e^-$}
\Text(50,60)[]{$e^-$}
\end{picture}
&&
\raisebox{10\unitlength}{\begin{tabular}{l}
$ -ie\gamma^{\mu}$\\
\end{tabular}}\\
\begin{picture}(100,60)(0,0)
\Photon(50,10)(10,10){4}{4}
\Vertex(50,10){2}
\ArrowLine(50,10)(90,10)
\ArrowLine(50,50)(50,10)
\Text(0,10)[]{$Z^{\mu}$}
\Text(100,10)[]{$e^-$}
\Text(50,60)[]{$e^-$}
\end{picture}
&&
\raisebox{30\unitlength}{\begin{tabular}{l}
$ie\gamma^{\mu}(a_e P_L + b_e P_R)$
\end{tabular}}\\
\begin{picture}(115,60)(0,0)
\Photon(50,10)(10,10){4}{4}
\Vertex(50,10){2}
\ArrowLine(50,10)(90,10)
\ArrowLine(50,50)(50,10)
\Text(0,10)[]{$\gamma^{\mu}$}
\Text(110,10)[]{$\chi_i^+$}
\Text(50,60)[]{$\chi_j^+$}
\end{picture}
&&
\raisebox{10\unitlength}{\begin{tabular}{l}
$ ie\gamma^{\mu}\delta_{ij}$\\
\end{tabular}}\\
\begin{picture}(100,60)(0,0)
\Photon(50,10)(10,10){4}{4}
\Vertex(50,10){2}
\ArrowLine(50,10)(90,10)
\ArrowLine(50,50)(50,10)
\Text(0,10)[]{$Z^{\mu}$}
\Text(100,10)[]{$\chi_i^+$}
\Text(50,60)[]{$\chi_j^+$}
\end{picture}
&&
\raisebox{30\unitlength}{\begin{tabular}{l}
$i\gamma^{\mu}\left(V_{LC}^{ij} P_L + V_{RC}^{ij} P_R\right)$
\end{tabular}}\\
\begin{picture}(100,60)(0,0)
\Photon(50,10)(10,10){4}{4}
\Vertex(50,10){2}
\ArrowLine(50,10)(90,10)
\ArrowLine(50,50)(50,10)
\Text(0,10)[]{$Z^{\mu}$}
\Text(100,10)[]{$\chi_i^0$}
\Text(50,60)[]{$\chi_j^0$}
\end{picture}
&&
\raisebox{30\unitlength}{\begin{tabular}{l}
$i\gamma^{\mu}\left(V_{N}^{ij} P_L - V_{N}^{ij\star} P_R\right)$
\end{tabular}}\\
\end{tabular}

\noindent where
\bea
V_{LC}^{ij} &=& - {e \over 2\sin\theta_W\cos\theta_W}
(Z_{1i}^{+*} Z_{1j}^{+} + \delta^{ij} \cos2\theta_W)\\
V_{RC}^{ij} &=& - {e \over 2\sin\theta_W\cos\theta_W}
(Z_{1i}^{-} Z_{1j}^{-*} + \delta^{ij} \cos2\theta_W)\\
V_{N}^{ij} &=&{e \over 2\sin\theta_W\cos\theta_W}
(Z_N^{4i*} Z_N^{4j} - Z_N^{3i*} Z_N^{3j})
\eea


\begin{thebibliography}{99}

\bibitem{WAGNER} M. Brhlik and G.L. Kane, {\em Phys. Lett.} {\bf
    B437}, (1998) 331, A. Pilaftsis, C.E.M. Wagner, {\em Nucl. Phys.}
  {\bf B553} (1999) 3-42; M. Carena, J. Ellis, A. Pilaftsis, C.E.M.
  Wagner, {\em Nucl. Phys.} {\bf B586} (2000) 92-140.

\bibitem{EDM_E_EXP} E. Commins {\em et al.}, {\em Phys. Rev.} {\bf
    A50} (1994) 2960; K. Abdullah {\em et al.}, {\em Phys. Rev. Lett.}
  {\bf 65} (1990) 234.

\bibitem{EDM_N_EXP} P. G. Harris {\em et al.}, {\em Phys. Rev. Lett.}
  {\bf 82}, (1999) 904.

\bibitem{STRONGCP} J. Ellis, S. Ferrara and D.V. Nanopoulos, {\em
    Phys. Lett.} {\bf 114B} (1982) 231; W. Buchm\"uller and D. Wyler,
  {\em Phys. Lett.} {\bf 121B} (1983) 321; J. Polchinski and M.B. Wise
  {\em Phys. Lett.} {\bf 125B} (1983) 393; J.M. Gerard {\em et al.},
  {\em Nucl. Phys.} {\bf B253} (1985) 93; P. Nath, {\em Phys. Rev.
    Lett.}  {\bf 66} (1991) 2565; Y. Kizuruki and N. Oshimo. {\em
    Phys. Rev.} {\bf D45} (1992) 1806; {\em Phys. Rev.} {\bf D46}
  (1992) 3025; R. Garisto, {\em Nucl.  Phys.} {\bf B419} (1994) 279.

\bibitem{OLIVE} T. Falk, K.A. Olive {\em Phys. Lett.} {\bf B439}
  (1998) 71; {\em Phys. Lett.} {\bf B375} (1996) 196.

\bibitem{NATH} T. Ibrahim and P. Nath, {\em Phys. Lett.} {\bf B418}
  (1998) 98; {\em Phys. Rev.} {\bf D57} (1998) 478; {\em Phys. Rev.}
  {\bf D58} (1998) 111301.

\bibitem{BARTL} A. Bartl, T. Gajdosik, W. Porod, P. Stockinger and H.
  Stremnitzer, {\em Phys. Rev.} {\bf D60} (1999) 073003.

\bibitem{KANE} M. Brhlik, G.J. Good and G.L. Kane, {\em Phys. Rev.}
  {\bf D59} (1999) 115004.

\bibitem{PRS} S. Pokorski, J. Rosiek and C. Savoy, {\em Nucl. Phys.}
  {\bf B570} (2000) 81-116.

\bibitem{BHRLIK} M. Brhlik, L. Everett, G.L. Kane, J. Lykken, {\em
    Phys. Rev. Lett.} {\bf 83} (1999) 2124-2127.

\bibitem{FCNC97} M. Misiak, J. Rosiek, S. Pokorski, {\em
    hep-ph}/9703442, published in A. Buras, M. Lindner (eds.), {\em
    Heavy flavours II}, pp. 795-828, World Scientific Publishing Co.,
  Singapore.

\bibitem{FEYRUL} J. Rosiek, {\em Phys. Rev.} {\bf D41}, (1990) 3464,
  {\em erratum hep-ph}/9511250.

\bibitem{SUSYGEN} N. Ghodbane {\sl et. al.}, {\em hep-ph}/9909499.
  
\bibitem{CHOI} 
G. Moortgat-Pick et al., Eur. Phys. J. C 7 (1999) 113; G. Moortgat-Pick et al.,Eur. Phys. J. C 9 (1999) 521; S. Y. Choi, M. Guchait, J. Kalinowski, P. M. Zerwas,  {\em Phys. Lett.} {\bf B479} (2000) 235-245; S. Y. Choi, A. Djouadi,
  M. Guchait, J. Kalinowski, H. S. Song, P. M. Zerwas, {\em Eur. Phys.    J.} {\bf C14} (2000) 535-546.

\bibitem{DELPHI} {\it Search for charginos in $e^+e^-$ interactions at
    $\sqrt{s} = 189$ GeV.} DELPHI collaboration {\em CERN-EP
    2000-008}, accepted for publication at Physics Letters.

\bibitem{moenig} K. M\"onig, \textit{Model independent limit of the
    Z-decays Width into Unknown particles}, DELPHI 97-174 PHYS 748.
  
\bibitem{MUNOZ} J.A. Casas, A. Lleyda, C. Munoz, {\em Nucl. Phys.}
  {\bf B471} (1996) 3-58.
   
\bibitem{KANE2} G. L. Kane, L. Wang, {\em Phys. Lett.} {\bf B488}
  (2000) 383-389.
  
\bibitem{DEGCHARG} DELPHI Collaboration ``Search for charginos nearly
  mass-degenerate with the lightest neutralino'', {\em Eur. Phys. J.}
  {\bf C11} (1999) 1; L3 Collaboration, {\em Phys. Lett.} {\bf B482}
  (2000) 31.
  
\bibitem{GUNION} C-H. Chen, M. Drees, J.F. Gunion, {\em Phys. Rev.}
  {\bf D55} (1997) 330-347; {\em erratum ibid} {\bf D60} (1999)
  039901.

\end{thebibliography}
\end{document}